\documentclass[12pt,a4paper]{article}

\usepackage{amsmath,amsthm,amssymb}
\usepackage{microtype}
\usepackage{xcolor}
\usepackage{natbib}
\usepackage[utf8]{inputenc}
\usepackage{csquotes}
\usepackage{multirow}
\usepackage{float}
\usepackage{mathtools}
\usepackage{graphicx}
\usepackage{longtable}
\usepackage[symbol]{footmisc}

\usepackage{tikz}

\graphicspath{ {./Grafiken/} }

\usepackage{booktabs,array}

\newcount\rowc

\makeatletter
\def\ttabular{%
\hbox\bgroup
\let\\\cr
\def\rulea{\ifnum\rowc=\@ne \hrule height 1.3pt \fi}
\def\ruleb{
\ifnum\rowc=1\hrule height 1.3pt \else
\ifnum\rowc=6\hrule height \heavyrulewidth 
   \else \hrule height \lightrulewidth\fi\fi}
\valign\bgroup
\global\rowc\@ne
\rulea
\hbox to 10em{\strut \hfill##\hfill}%
\ruleb
&&%
\global\advance\rowc\@ne
\hbox to 10em{\strut\hfill##\hfill}%
\ruleb
\cr}
\def\endttabular{%
\crcr\egroup\egroup}
\title{Fisher transformation based Confidence Intervals of Correlations in Fixed- and Random-Effects Meta-Analysis}
\author{Thilo Welz\footnote{Correspondence: thilo.welz@tu-dortmund.de}, Philipp Doebler, Markus Pauly}
\date{\today}

\DeclareMathOperator{\Cov}{Cov}
\DeclareMathOperator{\Var}{Var}
\DeclareMathOperator{\E}{E}

\begin{document}

\maketitle
\thispagestyle{empty}
\begin{abstract}
Meta-analyses of %Pearson 
correlation coefficients are an important technique to integrate results from many cross-sectional and longitudinal research designs. Uncertainty in pooled estimates is typically assessed with the help of confidence intervals, which can double as hypothesis tests for two-sided hypotheses about the underlying correlation. A standard approach to construct confidence intervals for the main effect is the Hedges-Olkin-Vevea Fisher-z (HOVz) approach, which is based on the Fisher-z transformation. Results from previous studies \citep{field2005meta, hafdahl2009meta}, however, indicate that in random-effects models the performance of the HOVz confidence interval can be unsatisfactory. To this end, we propose improvements of the HOVz approach, which are based on enhanced variance estimators for the main effect estimate. In order to study the coverage of the new confidence intervals in both fixed- and random-effects meta-analysis models, we perform an extensive simulation study, comparing them to established approaches. Data were generated via a truncated normal and beta distribution model. The results show that our newly proposed confidence intervals based on a Knapp-Hartung-type variance estimator or robust heteroscedasticity consistent sandwich estimators  in combination with the integral z-to-r transformation \citep{hafdahl2009improved} provide more accurate coverage than existing approaches in most scenarios, especially in the more appropriate beta distribution simulation model.
\end{abstract}


\newpage
\section{Introduction}
Quantifying the association of metric variables with the help of the Pearson correlation coefficient is a routine statistical technique to understand patterns of association. It is a basic ingredient of the data analysis of many cross-sectional and longitudinal designs, and is also indispensable for various psychometric and factor analytic techniques.  When several reports are available for comparable underlying populations, meta-analytic methods allow to pool the available evidence \citep{hedges1985statistical,hunter2004methods}, resulting in more stable and precise estimates. 

Systematic reviews based on meta-analyses of correlations are among the most cited in I/O-psychology, clinical psychology and educational psychology \citep[e.g.][each with several thousand citations]{barrick1991big,aldao2010emotion,sirin2005socioeconomic}, and the methodological monograph on pooling correlations of \citet{hunter2004methods} is approaching 10,000 citations on Google Scholar at the time of writing this article. In addition, pooled correlations are the basis for meta-analytic structural equation modeling \citep[e.g.,][]{cheung2015meta,jak2015meta}, and registered replication efforts pool correlations to re-assess findings of others \citep[e.g.,][]{open2015estimating}.

\subsection{The importance of confidence intervals for pooled correlations}
\citet{schulze2004meta} provides a comprehensive summary of fixed- and random-effects meta-analysis of correlations. The most well known approaches are based on Fisher's z-transformation \citep{hedges1985statistical,field2001meta, field2005meta, hafdahl2009meta} or on direct synthesis of correlations via the Hunter-Schmidt method \citep{hunter1994estimation, schulze2004meta}. Regardless of the method and the purpose of the meta-analysis, the point estimate of the correlation is to be accompanied by an estimate of its uncertainty, i.e., a standard error (SE) or a confidence interval (CI). Since the absolute value of a correlation is bounded by one, a CI might be asymmetric in this context, i.e., not centered around the point estimate. Also, CIs are often more useful than SEs, because a null hypothesis of the form $H_0:\rho = \rho_0$ can be rejected at level $\alpha$, if an $(1-\alpha)$-CI does not include $\rho_0$ (duality of hypothesis testing and CIs). A CI's coverage is ideally close to the nominal $(1-\alpha)$-level, e.g., a multi-center registered replication report does neither want to rely on an anti-conservative (too narrow) CI that is overly prone to erroneously rejecting previous research, nor on a conservative (too wide) CI lacking statistical power to refute overly optimistic point estimates. Despite methodological developments since the late 70s, the choice of a CI for a pooled correlation should be a careful one: Simulation experiments reported in this article reinforce the finding that CIs are too liberal when heterogeneity is present. The main objective of this paper is a systematic investigation of competing methods, especially when moderate or even substantial amounts of heterogeneity are present, promising refined meta-analytic methods for correlations, especially those based on the Fisher z-transformation. The remainder of the introduction reviews 
results for (z-transformation based) pooling, and briefly introduces relevant methods for variance estimation.

\subsection{Pooling (transformed) correlation coefficients}
A line of research summarized in \citet{hunter1994estimation} pools correlation coefficients on the original scale from $-1$ to $1$. One of the merits of the Hunter-Schmidt (HS) Methodology  is a clear rationale for artefact corrections, i.e., correlations are disattenuated for differences at the primary report level in reliability or variable range. While this part of the HS methodology is beyond the scope of the current paper, CIs originating from \citet{osburn1992note} are studied here as an HS-based reference method, which were also studied by \citet{field2005meta}.

Fisher's $z$-transformation (=areatangens hyperbolicus)  maps the open interval $(-1,1)$ to the real number line. Working with $z$ values of correlations avoids problems arising at the bounds and makes normality assumptions of some meta-analytic models more plausible \citep{hedges1985statistical}.  \citet{field2001meta} presents a systematic simulation study, and describes scenarios with a too liberal behavior of the HS methodology, but also reports problems with $z$-transformed pooled values. A simulation strategy is also at the core of \citet{field2005meta}, who places a special emphasis on heterogeneous settings. He finds similar point estimates for $z$-transformation based and HS pooling, with the CIs from the HS method too narrow in the small sample case. The simulation study of \citet{hafdahl2009meta} includes a comprehensive account of random-effects modeling and related sources of bias in point estimates. Focusing on point estimation, \citet{hafdahl2009meta} defend $z$-transformed pooling, but \citet{hafdahl2009improved} recommends the integral z-to-r transformation as a further improvement. In the spirit of  \citet{hafdahl2009meta}, the current paper focuses on variance estimators and resulting CIs, especially in the case of heterogeneity. 

\subsection{Estimating between study variance}
All CIs studied here are of the form $g(\hat\theta \pm \hat{\sigma}_{\hat\theta})$, for an appropriate back-transformation $g$ (which is not needed in the HS approach), a point estimator $\hat\theta$ and its SE estimator $\hat{\sigma}_{\hat\theta}$, which depends on the between-study variance estimation. The CI's quality will depend on an appropriate choice. In other words, especially when primary reports are heterogeneous and the underlying study-specific true correlations vary, good estimators of the between study variance are needed to obtain neither too wide nor too narrow CIs. 

The comprehensive study of \citet{veroniki2016methods} supports restricted maximum likelihood estimation (REML) as a default estimator of the between study variance. Since large values of the mean correlation cause REML convergence problems, the robust two-step  \citet{sidik2006robust} estimator is adopted here. Recently, \citet{welz2020simulation} showed that in the context of meta-regression, the Knapp-Hartung-adjustment \citep[][KH henceforth]{hartung1999alternative, hartung2001refined} aided (co-)variance estimation, motivating to include KH-type CIs in the subsequent comparison.

Less well known in the meta-analysis literature are bootstrap methods for variance estimation, which are not necessarily based on a parametric assumption for the random effects distribution. The \citet{wu1986jackknife} Wild Bootstrap (WBS) intended for heteroscedastic situations is evaluated here. Bootstrapping is complemented by Sandwich estimators \citep[heteroscedasticity consistent, HC;][]{white1980heteroskedasticity} that \citet{viechtbauer2015comparison} introduced in the field of meta-analysis. Recently, a wide range of HC estimators were calculated by \citet{welz2020simulation}, whose comparison also includes the more recent HC4 and HC5 estimators \citep{cribari2004leverage,cribari2007inference}. In sum, the following comparison includes a comprehensive collection of established and current variance estimators and resulting CIs.

In Section 2 we introduce the relevant models and procedures for meta-analyses of correlations with more technical detail, as well as our proposed refinements. In Section 3 we perform an extensive simulation study and present the results. An illustrative data example on the association of  conscientiousness \citep[in the sense of the NEO-PI-R;][]{costa1985neo, costa2008revised} and medication adherence \citep{molloy2013conscientiousness}
is presented in Section 4. We finally close with a discussion of our findings and give an outlook for future research.

\section{Meta-analyses of Pearson correlation coefficients}
\label{methods_normal}

For a bivariate metric random vector $(X,Y)$ with existing second moments the correlation coefficient $\varrho = \Cov(X,Y)/\sqrt{\Var(X)\Var(Y)}$ is usually estimated with the (Pearson) correlation coefficient

\begin{equation}
r = \frac{\sum\limits_{i=1}^n (x_i - \bar{x}) (y_i - \bar{y})}{\sqrt{\sum\limits_{i=1}^n (x_i - \bar{x})^2} \sqrt{\sum\limits_{i=1}^n (y_i - \bar{y})^2}},
\end{equation}

\noindent
where $(x_i,y_i), \ i=1,\ldots,n$, are independent observations of $(X,Y)$.

The Pearson correlation coefficient is asymptotically consistent, i.e., for large sample sizes, its value converges to the true $\varrho$. It is also invariant under linear transformations of the data. However, its distribution is difficult to describe analytically and it is not an unbiased estimator of $\varrho$   with an approximate bias of $\mathbb{E}(r - \varrho) \approx {-\tfrac{1}{2}\varrho(1-\varrho^2)}/{(n-1)}$ \citep{hotelling1953new}.

As correlation-based meta-analyses with $r$ as effect measure occur frequently in psychology and the social sciences we shortly recall the two standard models, cf. \cite{schwarzer2015meta}: the fixed- and random-effects model. The \textbf{fixed-effect} meta-analysis model is defined as
\begin{equation}
y_i = \mu + \varepsilon_i, \ i = 1,\ldots,K,
\end{equation}
\noindent
where $\mu$ denotes the common (true) effect, i.e., the (transformed) correlation in our case, $K$ the number of available primary reports, and $y_i$ the observed effect in the $i^{th}$ study. The model errors $\varepsilon_i$ are typically assumed to be normally distributed with $\varepsilon_i \overset{ind}\sim N(0,\sigma_i^2)$. In this model the only source of sampling error comes from \textit{within} the studies. The estimate of the main effect $\mu$ is then computed as a weighted mean via
\begin{equation}\label{iv_weighted_mean}
\hat{\mu} = \sum\limits_{i = 1}^K \frac{w_i}{w} y_i,
\end{equation}
\noindent
where $w \coloneqq \sum\limits_{i = 1}^K w_i$ and the study weights $w_i = \hat{\sigma}_i^{-2}$ are the reciprocals of the (estimated) sampling variances $\hat{\sigma}_i^2$. This is known as the \textit{inverse variance method}. The fixed-effect model typically underestimates the observed total variability because it does not account for between-study variability \citep{schwarzer2015meta}. However, it has the advantage of being able to pool observations, if individual patient data (IPD) are in fact available, allowing for greater flexibility in methodology in this scenario.

The \textbf{random-effects} model extends the fixed-effect model by incorporating a random-effect that accounts for between-study variability, such as differences in study population or execution. It is given by

\begin{equation}\label{eq:REmod}
\mu_i = \mu + u_i + \varepsilon_i, \ i = 1,\ldots,K,
\end{equation}

\noindent
where the random-effects $u_i$ are typically assumed to be independent and $N(0,\tau^2)$ distributed with between-study variance $\tau^2$ and $\varepsilon_i \overset{ind}{\sim} \mathcal{N}(0,\sigma_i^2)$. Furthermore, the random effects $(u_i)_i$ and the error terms $(\varepsilon_i)_i$ are jointly independent. Thus,  for $\tau^2 = 0$, the fixed-effect model is a special case of the random-effects model.
%In our simulation study we will at times relax these assumptions to $u_i \sim F$, where $F$ is a distribution with mean 0 and variance $\tau^2$, and $\varepsilon_i \sim G_i$, where $G_i$ is a distribution with mean 0 and variance $\sigma_i^2$.
The main effect is again estimated via the weighted mean $\hat{\mu}$ given in Equation \eqref{iv_weighted_mean} with study weights now defined as $w_i = (\hat{\sigma}_i^2 + \hat{\tau}^2)^{-1}$.

A plethora of approaches exist for estimating the heterogeneity variance $\tau^2$. Which estimator should be used has been discussed for a long time, without reaching a definitive conclusion. However, a consensus has been reached that the popular and easy to calculate DerSimonian-Laird estimator is not the best option. Authors such as \cite{veroniki2016methods} and \cite{langan2019comparison} have recommended to use iterative estimators for $\tau^2$. We therefore (initially) followed their suggestion and used the REML estimator. However, in some settings, such as large $\varrho$  values, the REML estimator had trouble converging, even after the usual remedies of utilizing step halving and/or increasing the maximum number of allowed iterations. We therefore opted to use the two-step estimator suggested by Sidik and Jonkman (SJ), which is defined by starting with a rough initial estimate of $\hat{\tau}^2_0 = \tfrac{1}{K}\sum_{i=1}^K (y_i - \bar{y})^2$ and is then updated via the expression
\begin{equation}
\hat{\tau}^2_{SJ} = \frac{1}{K-1} \sum_{i=1}^K w_i (y_i - \hat{\mu})^2,
\end{equation}

\noindent
where $w_i = \left( \frac{\hat{\tau}^2_0}{\hat{\sigma}_i^2 + \hat{\tau}^2_0} \right)^{-1}$ and $\hat{\mu} = \tfrac{\sum_{i=1}^K w_i y_i}{\sum_{i=1}^K w_i}$ \citep{sidik2005simple}. A comprehensive comparison of heterogeneity estimators for $\tau^2$ in the context of random-effects meta-analyses for correlations would be interesting but is beyond the scope of this paper. Before discussing different CIs for the common correlation $\mu$ within Model~\eqref{eq:REmod}, we take a short excursion on asymptotics for $r$ in the one group case.

%We decided on the non-iterative Hedges estimator for the sake of reducing computational costs, especially in combination with computationally expensive resampling approaches.} The Hedges Estimator, a non-iterative estimator of $\tau^2$,  is defined as
%
%\begin{equation}
%\hat{\tau}^2_{HE} \coloneqq \frac{\sum\limits_{i=1}^K (\varrho_i - \bar{\varrho})}{k - 1} - \frac{1}{K} \sum\limits_{i=1}^K \sigma_i^2,
%\end{equation}
%
%\noindent
%with $\bar{\varrho} = \frac{1}{K} \sum\limits_{i=1}^K \varrho_i$. The Hedges estimator has the advantage of being unbiased not only under exact knowledge of the sampling distributions $\sigma_i^2$ but also for unbiased estimates $\hat{\sigma}_i^2$, as well as being easy to compute. However, it is less efficient than some other estimators. \citep{viechtbauer2005bias} The Hedges estimator can unfortunately take on negative values, due to the involved differencing. Therefore we set $\hat{\tau}^2_{HE} \coloneqq \max\{\hat{\tau}^2_{HE},0\}$.

\subsection{Background: Asymptotic confidence intervals}\label{subsec:asymp}

Assuming bivariate normality of $(X,Y)$, $r$ is approximately $\mathcal{N}(\varrho,(1-\varrho^2)^2/n)$-distributed for large sample sizes $n$ \citep{lehmann2004elements}. Here, bivariate normality is a necessary assumption to obtain $(1-\varrho^2)^2$ in the asymptotic variance \citep{omelka2012testing}. Plugging in $r$, we obtain an approximate $(1-\alpha)$-CI of the form $r \pm u_{1-\alpha/2} {(1-r^2)}/{\sqrt{n}}$, where $u_{1-\alpha/2}$ denotes the $(1-\alpha/2)$-quantile of the standard normal distribution.

In fixed-effect meta-analyses, when IPD are available, this result can be used to construct a CI based on pooled data: Calculating $\hat{\varrho}_{pool}$ -- the pooled sample correlation coefficient -- we obtain an approximate CI for $\varrho$ by

\begin{equation}\label{poolCI}
\hat{\varrho}_{pool} \pm u_{1-\alpha/2} \frac{(1 - \hat{\varrho}_{pool}^2)}{\sqrt{N}},
\end{equation}

\noindent where $N \coloneqq \sum\limits_{i = 1}^K n_i$ is the pooled sample size. As this pooling of observations only makes sense if we assume that each study has the same underlying effect, this approach is not feasible in the case of a random-effects model, even if IPD were available. Anyhow, even under IPD and a fixed-effects model, this CI is sensitive to the normality assumption and the underlying sample size, as we demonstrate in Table \ref{mini_simu} for the case $K=1$. We simulated bivariate data from standard normal and standardized lognormal distributions\footnote[2]{Further details regarding the data generation can be found in the supplement.} with correlation $\varrho \in \{0.3,0.7\}$ and study size $n \in \{20,50,100\}$. Per setting we performed $N = 10,000$ simulation runs. For the lognormal data coverage is extremely poor in all cases, ranging from $53-80 \%$. For the normally distributed case coverage was somewhat low at $90\%$ for $n = 20$ but improved for larger sample sizes. This case study clearly illustrates that alternatives are needed, when the data cannot be assumed to stem from a normal distribution or sample sizes are small.

\begin{table}[ht]
\caption{Empirical coverage of the asymptotic confidence interval for $K=1$, study sizes $n \in \{20,50,100\}$ and correlations $\varrho \in \{0.3,0.7\}$.}
\label{mini_simu}
\centering
\begin{tabular}{ccccc}
\toprule
 Distribution & $\varrho$ & 20 & 50 & 100 \\ 
\midrule
\multirow{2}{*}{normal} & 0.3 & 0.90 & 0.93 & 0.94\\
& 0.7 & 0.90 & 0.92 & 0.94\\
\midrule
\multirow{2}{*}{lognormal} & 0.3 & 0.79 & 0.80 & 0.79\\
 & 0.7 & 0.63 & 0.57 & 0.53\\
\bottomrule
\end{tabular}
\end{table}

After this short excursion we turn back to Model~\eqref{eq:REmod} and CIs for  $\varrho$.

\subsection{The Hunter-Schmidt approach}

The aggregation of correlations in the Hunter-Schmidt approach is done by sample size weighting:
\begin{equation}
r_{HS} = \frac{\sum_{i=1}^K n_i r_i}{\sum_{i=1}^K n_i}.
\end{equation}

Several formulae have been recommended for estimating the sampling variance of this mean effect size estimate. We opted for a suggestion by \cite{osburn1992note}:

\begin{equation}
\hat{\sigma}_{HS}^2 = \frac{1}{K} \left( \frac{\sum_{i=1}^K n_i (r_i - r_{HS})^2}{\sum_{i=1}^K n_i} \right),
\end{equation}

\noindent
which is supposed to perform reasonably well in both heterogeneous and homogeneous settings \citep[][]{schulze2004meta}. In the simulation study we will investigate, whether this is in fact the case for the resulting CI: $r_{HS} \pm u_{1-\alpha/2}\hat{\sigma}_{HS}$.

\subsection{Confidence Intervals based on the Fisher-z transformation}

A disadvantage of the asymptotic confidence interval (\ref{poolCI}) is that the variance of the limit distribution depends on the unknown correlation $\varrho$. This motivates a variance stabilizing transformation. A popular choice for correlation coefficients is the \textbf{Fisher-z transformation}  \citep{fisher1915frequency},%. The Fisher-z transformation of the population correlation coefficient $\varrho$ is defined as
\begin{equation}
\rho \mapsto z = \frac{1}{2} \ln \left(\frac{1+\varrho}{1-\varrho} \right) = \text{atanh}(\varrho).
\end{equation}
The corresponding inverse Fisher transformation is $z \mapsto \tanh(z) = (\exp(2z)-1)/(\exp(2z) + 1)$.

The variance stabilizing property of the Fisher transformation follows from the $\delta$-method \citep{lehmann2004elements}, i.e., if $\sqrt{n}(r - \varrho) \overset{d}{\longrightarrow} \mathcal{N}\left( 0,(1 - \varrho^2)^2 \right)$
%can be verified via : Under normality we have $\sqrt{n} \big(r - \varrho \big) \overset{d}{\longrightarrow} \mathcal{N} \Big(0,(1-\varrho^2)^2 \Big)$, with the $\delta$-method it follows that 
then $\sqrt{n}(\hat{z}-z) = \sqrt{n} \big( \text{atanh}(r) - \text{atanh}(\varrho) \big) \overset{d}{\longrightarrow} \mathcal{N}(0,1).$ Following \cite{schulze2004meta}, it is reasonable to substitute $\sqrt{n}$ by $\sqrt{n-3}$, i.e., to approximate the distribution of $\hat{z}$ by $\mathcal{N} \left( \text{atanh}(r),\frac{1}{n-3} \right)$ -- still assuming bivariate normality. 
%As the inverse of the Fisher-$z$ transformation is given by
%\begin{equation*}
%\varrho = \frac{\exp(2z) -1}{\exp(2z) + 1} = \tanh(z),
%\end{equation*}
%\noindent
Thus, a single group approximate $(1-\alpha)$-CI can be constructed via
%\begin{equation*}
$\tanh \big( \hat{z} \pm {u_{1-\alpha/2}}/{\sqrt{N-3}} \big).$
%\end{equation*}
%The Fisher-z transformation is rather inaccurate for large $\varrho$ and small $N$. \citep{schulze2004meta} Therefore, more sophisticated approaches are needed in such scenarios.

In the random-effects model \eqref{eq:REmod}, the z-transformation may also be used to construct a CI for the common correlation $\varrho$. Here, the idea is again to use inverse variance weights to define

\begin{equation}
\bar{z} = \frac{\sum\limits_{i = 1}^K \left(\frac{1}{n_i - 3} + \hat{\tau}^2\right)^{-1} z_i}{\sum\limits_{i = 1}^K \left(\frac{1}{n_i - 3} + \hat{\tau}^2\right)^{-1}},
\end{equation}
where $z_i=\text{atanh}(r_i)$. 
%\noindent which is then transformed back via $\bar{r} = \tanh(\bar{z})$. 
A rough estimate of the variance of $\bar{z}$ is given by $\big( \sum_{i=1}^K w_i \big)^{-1}$. In the fixed-effect casel with $\tau^2 = 0$ this yields the variance estimate $\Big( \sum_{i=1}^K (n_i -3) \Big)^{-1} = \big(N - 3K \big)^{-1}$. Then $\bar{z} \sqrt{N - 3K}$ approximately follows a standard normal distribution and an approximate $(1 - \alpha)$-CI is given by
%\begin{equation}
$\tanh( \bar{z} \pm {u_{1-\alpha/2}}/{\sqrt{N - 3K}}).$ 
%\end{equation}
Proceeding similarly in the random-effects model \eqref{eq:REmod}, one obtains the 
HOVz CI  (\textit{Hedges-Olkin-Vevea Fisher-z})
%approximate $(1-\alpha)$ CI
\begin{equation}
\tanh \Big( \bar{z} \pm {u_{1-\alpha/2}}/{{\big( \sum_{i=1}^K w_i \big)^{1/2}}}\Big),
\end{equation}
\noindent
with $w_i = (\frac{1}{n_i - 3} + \hat{\tau}^2)^{-1}$  \citep{hedges1985statistical,hedges1998fixed,hafdahl2009meta}. 

\subsubsection{Knapp-Hartung-type CI}
The above approximation of the variance of $\bar{z}$ via $\left( \sum_{i=1}^K w_i \right) ^{-1}$ can be rather inaccurate, especially in random-effects models. Although this is the exact variance of $\bar{z}$ when the weights are chosen perfectly as $w_i = (\sigma_i^2 + \tau^2)^{-1}$, this variance estimate does not protect against (potentially substantial) errors in estimating $\hat{\sigma}_i^2$ and $\hat{\tau}^2$  \citep{sidik2006robust}. Therefore, we propose an improved CI based on the Knapp-Hartung method \citep[KH][]{hartung2001refined}. KH proposed the following variance estimator for the estimate $\hat{\mu}$ of the main effect $\mu$ in a random-effects meta-analysis:

\begin{equation}\label{knapp-hartung}
\hat{\sigma}_{KH}^2 = \widehat{\text{Var}}_{KH}(\hat{\mu}) = \frac{1}{K-1} \sum\limits_{i=1}^K \frac{w_i}{w} \left(\hat{\mu}_i - \hat{\mu} \right) ^2,
\end{equation}
\noindent
where again $w = \sum_{i=1}^K w_i$. 
\cite{hartung1999alternative} showed that if $\hat{\mu}$ is normally distributed, then ${(\hat{\mu} - \mu)}/{\hat{\sigma}_{KH}}$ follows a $t$-distribution with $K-1$ degrees of freedom. Therefore an approximate $(1-\alpha)$-CI for $\mu$ is given by
\begin{equation}\label{KH_CI}
\tanh \big( \bar{z} \pm  t_{K-1,1-\alpha/2} \cdot \hat{\sigma}_{KH}\big),
\end{equation}
\noindent
where $t_{K-1,1-\alpha/2}$ is the $1-\alpha/2$ quantile of the $t$-distribution with $K-1$ degrees of freedom. %We compare these and the following approaches from section \ref{wildbs} in our simulation study in section \ref{simulation}. 
Because of the approximate normal distribution of z-transformed correlations, the CI (\ref{KH_CI}) seems justified.
Various authors have highlighted the favorable performance of the KH approach compared to alternative meta-analytic methods \citep{inthout2014hartung,viechtbauer2015comparison,welz2020simulation}. %Thus we believe it is reasonable to expect the KH method to also be competitive in the settings discussed in this paper.
Analogously to (\ref{KH_CI}), we can construct further CIs by using other variance estimation procedures for $\text{Var}(\hat{\mu})$.

\subsubsection{Wild Bootstrap Approach}
\label{wildbs}

Another possibility of estimating the variance of $\bar{z}$ is through bootstrapping. Bootstrapping belongs to the class of resampling methods. It allows the estimation of the sampling distribution of most statistics using random sampling methods. The wild bootstrap is a subtype of bootstrapping that is applicable in models, which exhibit heteroscedasticity. Roughly speaking, the idea of the wild bootstrap approach is to resample the response variables based on the residuals. The idea was originally proposed by \cite{wu1986jackknife} for regression analysis.

We now propose a confidence interval for $\varrho$ based on a (data-dependent) \textbf{wild-bootstrap approach} (WBS) combined with the z-transformation. The idea works as follows: We assume a random-effects meta-analysis model with Pearson's correlation coefficient as the effect estimate (and $K > 3$ studies). Given the estimated study level correlation coefficients $r_i, \ i=1,\ldots,K$, we transform these using z-transformation to $\hat{z}_i, \ i=1,\ldots,K$, and estimate $z = \text{atanh}(\varrho)$ via $\hat{z} = \sum_i \frac{w_i}{w} \hat{z}_i$, where again $w_i = (\hat{\sigma}_i + \hat{\tau}^2)^{-1}$ with $\hat{\sigma}_i^2 = \frac{1}{n_i - 3}$ and $w = \sum_i w_i$. Here, $\hat{\tau}^2$ may be any consistent estimator of the between-study heterogeneity $\tau^2$, where we have chosen the SJ estimator. We then calculate the estimated residuals $\hat{\varepsilon}_i = \hat{z} - \hat{z}_i$ and use these to generate $B$ new sets of study-level effects $\hat{z}_{1b}^*,\ldots,\hat{z}_{Kb}^*, \ b = 1,\ldots,B$. Typical choices for $B$ are 1,000 or 5,000. The new study-level effects are generated via

\begin{equation}
\hat{z}_{ib}^* := \hat{z}_i + \hat{\varepsilon}_i \cdot v_i,
\end{equation}

\noindent
where $v_i \sim \mathcal{N}(0,\gamma)$. The usual choice of variance in a wild bootstrap is $\gamma = 1$. However, we propose a data dependent choice of either $\gamma_K = \frac{K-1}{K-3}$ or $\gamma_K = \frac{K-2}{K-3}$. These choices are based on simulation results, which will be discussed in detail in Section \ref{simulation}. We will later refer to these approaches as WBS1, WBS2 and WBS3 respectively. The corresponding values for $\gamma$ are 1, $(K-1)/(K-3)$ and $(K-2)/(K-3)$. This allows us to generate $B$ new estimates of the main effect $z$ by calculating

\begin{equation}
\hat{z}_b^* = \frac{\sum_{i=1}^K w_{ib}^* \hat{z}_{ib}^*}{\sum_{i=1}^K w_{ib}^*},
\end{equation}

\noindent
with $w_{ib}^* \equiv w_i$. %An alternative approach would have been to set $w_{ib}^* = (\frac{1}{n_i - 3} + \hat{\tau}_b^{*2})^{-1}$, where $\hat{\tau}_b^{*2}$ is calculated based on the new set of estimates $\hat{z}_{1b}^*,\ldots,\hat{z}_{Kb}^*$, using the same type of heterogeneity estimator as for the original data. However, we only considered the former approach in our simulation study.

\begin{figure}[H]
\centering
\includegraphics[width=\textwidth]{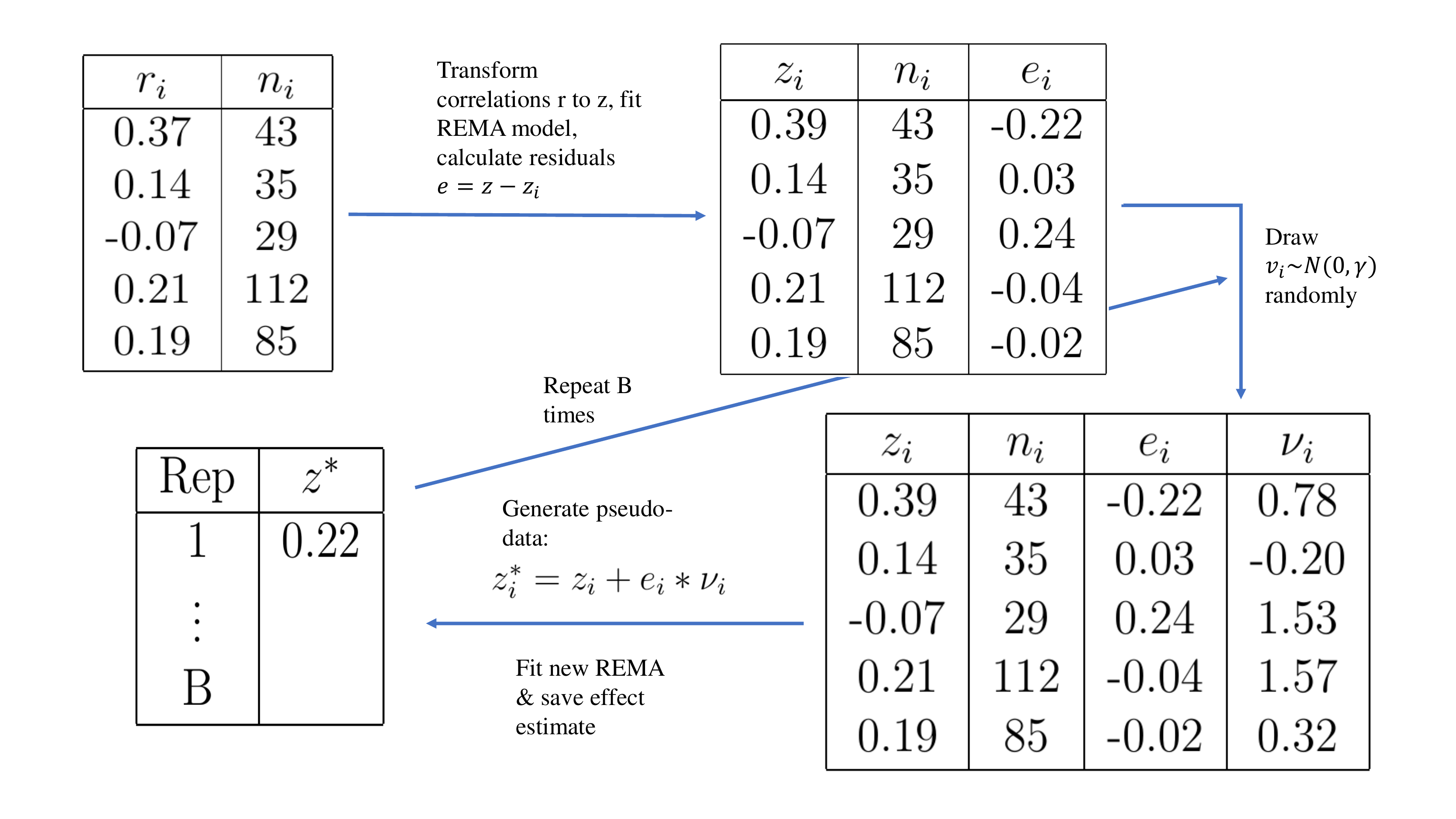}
\caption{Visual illustration of the Wild Bootstrap Procedure for generating $B$ bootstrap samples of the main effect estimate on the z-scale}
\label{WBS_illustration}
\end{figure}

We then estimate the variance of $\hat{z}$ via the empirical variance of $\hat{z}_1^*,\ldots,\hat{z}_B^*$, $\sigma_z^{*2} := \frac{1}{B-1} \sum\limits_{i=1}^B \left( \hat{z}_i^* - \bar{z}^* \right) ^2$ with $\bar{z}^* = \frac{1}{B} \sum_{i=1}^B \hat{z}_i^*$. It is now possible to construct a CI for $z$ as in Equation \eqref{KH_CI} but with this new variance estimate of $\bar{z}$. The CI is back-transformed via the inverse Fisher transformation to obtain a CI for the common correlation $\varrho$, given by
\begin{equation}\label{Wild_CI}
\tanh \Big( \hat{z} \pm \hat{\sigma}_z^* \cdot  t_{K-1,1-\alpha/2} \Big).
\end{equation}
%\noindent
%where $t_{K-1,1-\alpha/2}$ is again the $(1-\alpha/2)$-quantile of the $t_{K-1}$ distribution.
Figure \ref{WBS_illustration}  provides a visual illustration of the wild bootstrap procedure discussed above.

\subsubsection{HC-type variance estimators}

Last but not least, we employ \textbf{heteroscedasticity consistent (HC)} variance estimators  \citet[sandwich estimators][]{white1980heteroskedasticity}. Different forms ($HC_0-HC_5$) are in use for  linear models \citep{rosopa2013managing}. The motivation for the robust HC variance estimators is that in a linear regression setting the usual variance estimate is unbiased when unit level errors are independent and identically distributed. However, when the unit level variances are unequal, this approach can be biased. If we apply this to the meta-analysis context, the study level variances are almost always unequal due to varying sample sizes. Therefore, it makes sense to consider variance estimators that are unbiased even when the variances of the unit (study) level variances are different.

The extension of HC estimators to the meta-analysis context can be found in \cite{viechtbauer2015comparison} for $HC_0-HC_1$ and in \cite{welz2020simulation} for the remaining $HC_2-HC_5$. Statistical tests based on these robust estimators have been shown to perform well, especially those of types $HC_3$ and $HC_4$. In the special case of a random-effects meta-analysis they are defined as \citep[see the supplementary material of][for details]{welz2020simulation} 
\begin{align*}
\hat{\sigma}^2_{HC_3} & = \frac{1}{ \big( \sum_{i=1}^K w_i \big) ^2} \sum_{j=1}^K w_j^2 \hat{\varepsilon}^2_j(1-x_{jj})^{-2},\\
\hat{\sigma}^2_{HC_4} & = \frac{1}{ \big( \sum_{i=1}^K w_i \big) ^2} \sum_{j=1}^K w_j^2 \hat{\varepsilon}^2_j(1-x_{jj})^{-\delta_j}, \ \delta_j = \min \left\{4,\frac{x_{jj}}{\bar{x}} \right\}
%\hat{\sigma}^2_{HC_5} & = \frac{1}{\big( \sum_{i=1}^K w_i \big) ^2} \sum_{j=1}^K w_j^2 \hat{\varepsilon}^2_j(1-x_{jj})^{-\alpha_j}, \ \alpha_j = \min \left\{ \frac{x_{jj}}{\bar{x}},\max\{4,\frac{\eta x_{max}}{\bar{x}}\} \right\}
\end{align*}
with $\hat{\varepsilon}_j = \hat{z}_j - \hat{z}$, $x_{jj} = \frac{w_j}{\sum_{i=1}^K w_i}$ and $\bar{x} = \frac{1}{K} \sum\limits_{i=1}^K x_{ii}$. %Furthermore, $x_{max} = \max\{x_1,\ldots,x_K\}$ and $\eta$ is a tuning parameter that we set equal to 0.7 based on recommendations for regression models in the literature \citep{cribari2007inference}.
Plugging them into Equation \eqref{KH_CI} leads to the confidence intervals
\begin{equation}\label{HC_CI}
\tanh \Big( \hat{z} \pm \hat{\sigma}_{HC_j} \cdot  t_{K-1,1-\alpha/2} \Big), \ j = 3,4.
\end{equation}

\subsubsection{Integral z-to-r transformation}
\label{integral_ztor}
There is a fundamental problem with back-transforming CIs on z-scale using the inverse Fisher transformation $\tanh$: Consider a random variable $\xi \sim \mathcal{N}(\text{artanh}(\varrho),\sigma^2)$ with some variance $\sigma^2 > 0$and $\rho\neq 0$. Then $\varrho = \text{tanh}(\mathbb{E}(\xi)) \neq \mathbb{E}(\text{tanh}(\xi))$ by Jensen's inequality. This means the back-transformation introduces an additional bias. A remedy was proposed by \cite{hafdahl2009improved}, who suggested to instead backtransform from the z-scale using an integral z-to-r transformation. This transformation is the expected value of $\text{tanh}(z)$, where $z \sim \mathcal{N}(\mu_z,\tau^2_z)$, i.e.,

\begin{equation}
\psi(\mu_z \mid \tau_z^2) = \int_{-\infty}^{\infty} \text{tanh}(t)f(t \mid \mu_z,\tau_z^2)dt,
\end{equation}

\noindent
where $f$ is the density of $z$. In practice we apply this transformation to the lower and upper confidence limits on the z-scale, plugging in the estimates $\hat{z}$ and $\hat{\tau}_z^2$. For example, for the KH-based CI (\ref{KH_CI}) with z-scale confidence bounds $\ell=\bar{z}-t_{K-1,1-\alpha/2}\cdot \hat{\sigma}_{KH}$ and $u=\bar{z}+t_{K-1,1-\alpha/2}\cdot \hat{\sigma}_{KH}$, with an estimated heterogeneity $\hat{\tau}^2_z$ (on the z-scale), the CI is given by
\begin{equation*}
\left( \psi(\ell \mid \hat{\tau}^2_z),\psi(u \mid \hat{\tau}^2_z) \right).
\end{equation*}

If the true distribution of $\hat{z}$ is well approximated by a normal distribution and $\hat{\tau}^2_z$ is a good estimate of the heterogeneity variance (on the z-scale), $\psi$ should improve the CIs as compared to simply back-transformation with $\tanh$  \citep{hafdahl2009improved}. Following this argument, we also suggest using $\psi$ instead of $\tanh$. We calculate the integral with Simpson's rule \citep{suli2003introduction}, which is a method for the numerical approximation of definite integrals. 150 subintervals over $\hat{z} \pm 5\cdot \hat{\tau}_{SJ}$ were used, following \cite{hafdahl2009improved}. Note that the HOVz CI is implemented in its original formulation, using $\tanh$.

\section{Simulation Study}
\label{simulation}

We have suggested several new CIs for the mean correlation $\varrho$, all based on the z-transformation, applicable in both, fixed- and random-effects models. In order to investigate their properties (especially coverage of $\rho$), we perform extensive Monte Carlo simulations. We focus on comparing the coverage of our newly suggested CIs with existing methods.
% There already exist several suggestions on how to construct CIs in random-effects meta-analyses of correlations. Examples are the approaches by Hedges and Olkin, Rosenthal and Rubin, various versions of Hunter and Schmidt or the approach introduced by Olkin and Pratt, see \cite{schulze2004meta} for a comprehensive summary.
%We have chosen to compare our approaches with the most commonly used HOVz and Hunter-Schmidt (HS) approaches. Hereby we apply the integral z-to-r transformation suggested in \cite{hafdahl2009improved} for the newly proposed confidence intervals. This transformation was discussed in Subsection \ref{integral_ztor}. 

\subsection{Simulation study design} 
The Pearson correlation coefficient is constrained to the interval $[-1,1]$. The typical random-effects model $\mu_i = \mu + u_i + \varepsilon_i$, assuming a normal distribution for the random effect $u_i \sim \mathcal{N}(0,\tau^2)$ and error term $\varepsilon_i \sim \mathcal{N}(0,\sigma_i^2)$ needs to be adjusted, since values outside of $[-1,1]$ could result when sampling without any modification.

\noindent{\bf Model 1:} As a first option for generating the (true) study-level correlations, we consider a truncated normal distribution $\varrho_i \sim \mathcal{N}(\varrho,\tau^2)$: Sampling of $\varrho_i$ is repeated until a sample lies within the interval $[-0.999,0.999]$. %This process is repeated until $\mu_i$ lies in the specified interval. 
This type of truncated normal distribution model was also used in \cite{hafdahl2009meta} and \cite{field2005meta}. A problem with this modeling approach is that the expected value of the resulting truncated normal distribution is in general not equal to $\varrho$: For a random variable $X$ stemming from a truncated normal distribution with mean $\mu$ and variance $\sigma^2$ with lower bound $a$ and upper bound $b$,
% i.e., $X \sim T \mathcal{N}(\mu,\sigma^2,a,b)$, 
it holds that \citep{johnson1994continuous}
$$\mathbb{E}(X) = \mu + \sigma \frac{\phi(\Delta_1) - \phi(\Delta_2)}{\delta},$$
where $\Delta_1 = (a - \mu)/\sigma$, $\Delta_2 = (b - \mu)/\sigma$ and $\delta = \Phi(\Delta_2) - \Phi(\Delta_1)$. Here $\phi(\cdot)$ is the probability density function of the standard normal distribution and $\Phi(\cdot)$ its cumulative distribution function. Figure \ref{biasplot} in the supplement shows the bias in our setting with $a=-0.999$ and $b=0.999$. The bias is equal to $\sigma {(\phi(\Delta_1) - \phi(\Delta_2))}/{\delta}$. In addition to generating a biased effect, the truncation also leads to a reduction of the overall variance, which is smaller than $\tau^2$.

\noindent{\bf Model 2:} We therefore studied a second model, in which we generate the (true) study level effects $\varrho_i$ from transformed beta distributions: $Y_i = 2 (X_i - 0.5)$ with $X_i \sim Beta(\alpha, \beta)$ for studies $i = 1,\ldots,K$. The idea is to choose the respective shape parameters $\alpha, \beta$ such that the following equalities hold:

\begin{align*}
\E(Y_i) & = 2 \cdot \left( \frac{\alpha}{\alpha + \beta} - 0.5 \right) \stackrel{!}{=} \varrho, \\
\Var(Y_i) & = \frac{4 \alpha \beta}{(\alpha + \beta)^2 (\alpha + \beta + 1)}  \stackrel{!}{=} \tau^2.
\end{align*}

\noindent
The solution to the system of equations above is:

\begin{align*}
\alpha & = \frac{(1 - \varrho) (1 + \varrho) - \tau^2}{\tau^2} \cdot \left( \frac{1+\varrho}{2} \right),\\
\beta & = \left( \frac{1 - \varrho}{1 + \varrho} \right) \alpha.
\end{align*}

In this second simulation scenario we also truncate the sampling distribution of the correlation coefficients to $[-0.999,0.999]$, but values outside of this interval are considerably rarer. The second model has the advantages that the expected value and variance are approximately correct, unlike in the first (truncated) model. A disadvantage is that for extreme $\tau^2$ values, the above solution for $\alpha$ (and thus $\beta$) may become negative, which is undefined for parameters of a beta distribution. However, this was not a concern for the parameters considered in our simulation study and only occurs in more extreme scenarios.

\noindent{\bf Parameter choices.} In order to get a broad overview of the performance of all methods, we simulated various configurations of population correlation coefficient, heterogeneity, sample size and number of studies. Here we chose the correlations $\varrho \in \{0,0.1,0.3,0.5,0.6,0.7,0.8,0.9\}$ and heterogeneity $\tau \in \{0, 0.16, 0.4\}$. Moreover, we considered small to large number of $K \in \{5,10,20,40\}$ studies with different study sizes: For $K=5$, we considered $\vec n = (15,16,19,23,27)$ as vector of 'small' study sizes and $4 \cdot \vec n$ for larger study sizes, corresponding to an average study size $(\bar{n})$ of $20$ and $80$ subjects, respectively. For all other choices of $K$ we proceeded similarly, stacking copies $\vec n$ behind each other, e.g., the sample size vectors $(\vec n, \vec n)$ and $4\cdot (\vec n, \vec n)$ for $K=10$. Additionally, we considered two special scenarios: The case of few and heterogeneous studies, with study size vector $(23, 19, 250, 330, 29)$ and the case of many large studies, with study size vector $(\vec n^*, \vec n^*) $ with $\vec n^* = (210, 240, 350, 220, 290, 280, 340, 400, 380, 290)$. The latter case corresponds to $K = 20$ studies with an average of $300$ study subjects. 

Thus, in total we simulated $8 (\varrho) \times 3 (\tau^2) \times 10 (K, \text{study size vector}) \times 2$ (Model) $= 480$ different scenarios for each type of confidence interval discussed in this paper.
% plus the special setting of the data example from Section \ref{DataExampleSec}. 
For each scenario we performed $N =  10,000$ simulation runs, where for the WBS CI each run was based upon $B=1,000$ bootstrap replications. The primary focus was on comparing empirical coverage with nominal coverage being $1 - \alpha = 0.95$. For 10,000 iterations, the Monte Carlo standard error of the simulated coverage will be approximately $\sqrt{\tfrac{.95 \times .05}{10000}} \approx 0.218\%$, using the formula provided in the recent work on simulation studies by \citet{morris2019using}.

\subsection{Results}

For ease of presentation, we aggregated the multiple simulation settings with regard to number and size of studies. The graphics therefore display the mean observed coverage for each confidence interval type and true main effect $\varrho$. Results are separated by heterogeneity $\tau^2$ and simulation design. The latter refers to the truncated normal-distribution approach and the transformed beta-distribution approach respectively. More detailed simulation results for all considered settings are given in the supplement.

\subsubsection{Coverage}

\begin{figure}[H]
\centering
\input{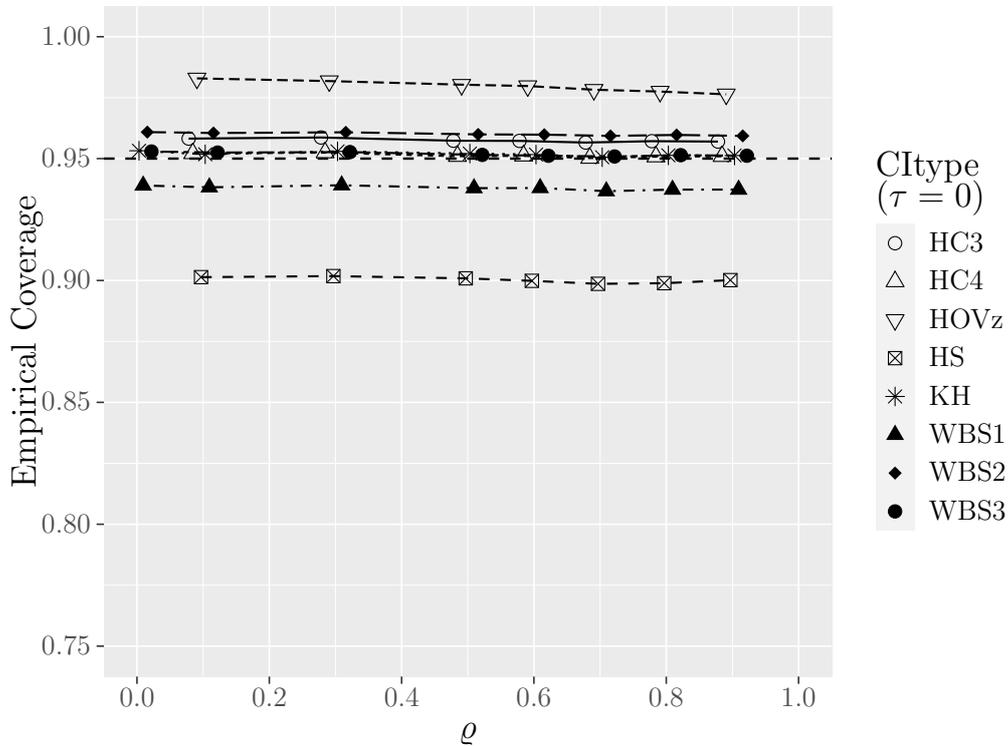}
\caption{Mean Coverage for truncated normal distribution model with $\tau = 0$, aggregated across all number of studies and study size settings}
\label{normal_0}
\end{figure}

We first discuss the results based on the truncated normal distribution (Model 1).  
In the case of no heterogeneity (fixed-effect model), Figure~\ref{normal_0} shows that the new methods control the nominal coverage of $95\%$ well. Only the first wild bootstrap (WBS1) CI exhibits a liberal behaviour, yielding empirical coverage of approximately $93.5\%$. The Hunter-Schmidt approach (HS) only provides $90\%$ coverage and HOVz was slightly conservative with (mean) coverage of around $97-98\%$. Moreover, in the fixed-effect model the value of $\varrho$ did not affect any of the methods.

\begin{figure}[H]
\centering
\input{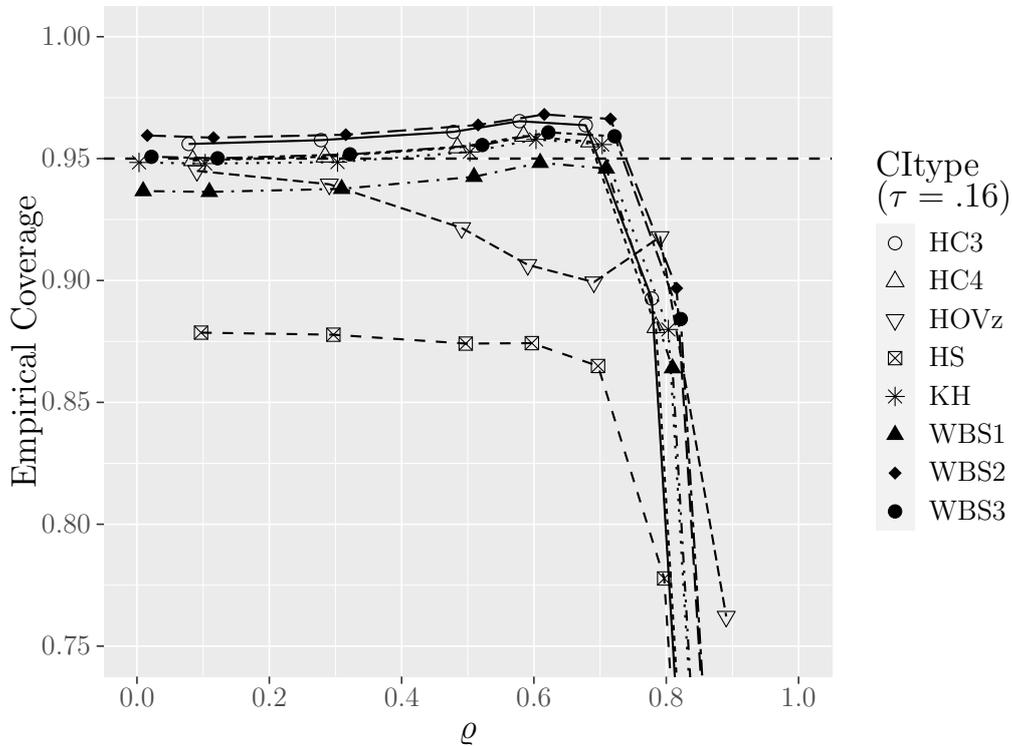}
\caption{Mean Coverage for truncated normal distribution model with $\tau = 0.16$, aggregated across all number of studies and study size settings}
\label{normal_16}
\end{figure}

In the truncated-normal setup with moderate heterogeneity of $\tau = 0.16$ in Figure \ref{normal_16}, several things change: First, there is a strong drop-off in coverage for larger correlations $\varrho \geq 0.8$. For HS this drop-off occurs earlier for $\varrho \geq 0.7$. Second, for $\varrho \leq 0.7$, HS is even more liberal than for $\tau=0$ with coverage around $87.5\%$. Additionally, HOVz is no longer conservative but becomes more liberal than WBS1 with estimated coverage probabilities around $90-94\%$ for $\varrho \leq 0.7$. For all new methods a slight decrease in coverage can be observed for increasing values of $\varrho$ from $0$ to $0.7$. Moreover, there is a slight uptick at $\varrho = 0.8$ for HOVz, followed by a substantial drop-off. Overall the WBS3, $HC_3$, $HC_4$ and KH CIs show the best control of nominal coverage in this setting.

%\begin{figure}[H]
%%\includegraphics[width=\textwidth]{beta_0.pdf}
%\centering
%\input{Grafiken/beta_0.tex}
%\caption{Mean Coverage for transformed beta distribution model with $\tau = 0$, aggregated across all settings for number of studies and study size}
%\label{beta_0}
%\end{figure}

\begin{figure}[H]
\centering
\input{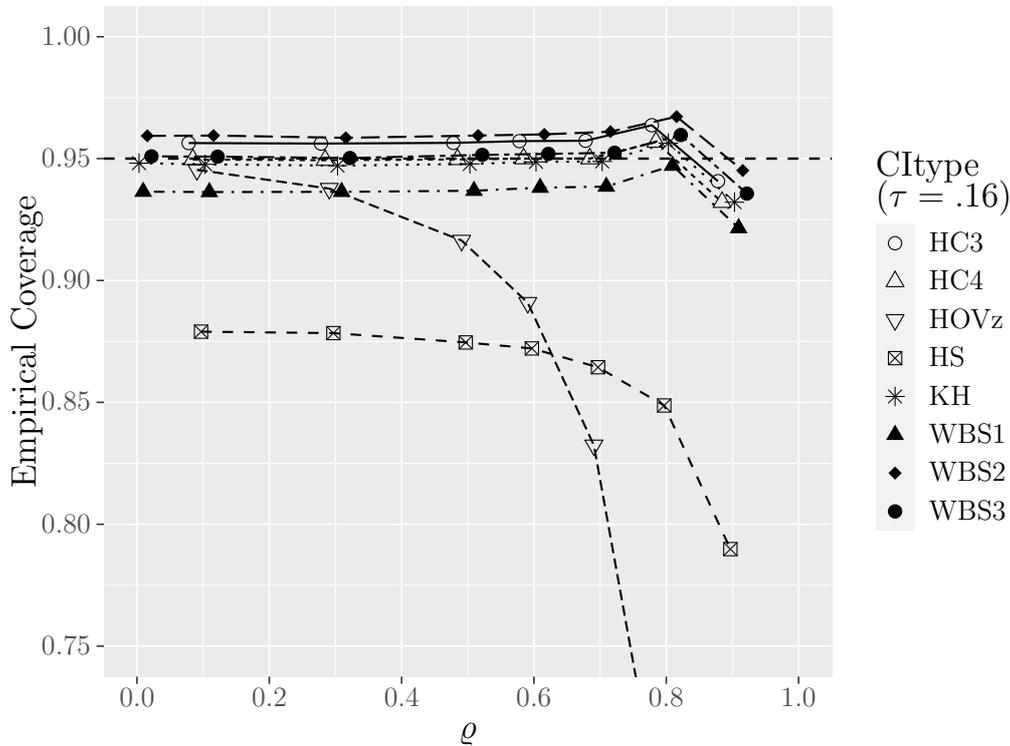}
\caption{Mean Coverage for transformed beta distribution model with $\tau = 0.16$, aggregated across all number of studies and study size settings}
\label{beta_16}
\end{figure}

We now consider Model 2 with a transformed beta distribution model. In the fixed effects case ($\tau^2=0$) the two models are equivalent so we obtain the same coverage as in Figure \ref{normal_0}. For moderate heterogeneity ($\tau = 0.16$, cf. Figure \ref{beta_16}), our newly proposed methods clearly outperform HOVz and HS, with a good control of nominal coverage. Only for $\varrho=0.9$ their coverage is slightly liberal. WBS1 performs just slightly worse than the other new CIs. The observed coverage for HS lies at $\approx 86-88\%$ for $\varrho \leq 0.7$ and drops to just below $80\%$ for $\varrho = 0.9$. For $\varrho > 0.6$ the HOVz CI is even worse with values dropping (substantially) below 75\%.

For ease of presentation, the results for the case of extreme heterogeneity with $\tau = 0.4$ are given in the supplement. Here, we only summarize important points from the Figures \ref{normal_4}--\ref{beta_4}. In the truncated normal distribution model we observe that HS again has unsatisfactory coverage, compared with the other approaches. For our new CIs based on the Fisher transformation, for small $K$, coverage is approximately correct for $\varrho \leq 0.6$ and then drops off considerably. HOVz is slightly liberal with coverage around $90\%$ for $\varrho \leq 0.6$ and then drops off strongly. This holds for both smaller and larger studies with $\bar{n} \in \{20,80\}$ respectively. For an increasing number of studies $K$, HOVz remains largely unchanged, whereas coverage of the new methods gets progressively worse (i.e. the drop-off in coverage occurs earlier for an increasing number of studies). For $K = 40$ the new CIs only have correct coverage for $\varrho \leq 0.3$. In the case of the beta distribution model with $\tau = 0.4$ the new CIs provide correct coverage for $\varrho \leq 0.7$ in all scenarios, dropping off after this threshold. HOVz is very inadequate, with coverage getting progressively worse for increasing $K$. HOVz only has correct coverage for simultaneously $\varrho \leq 0.1$ and large $K$. For $K=5$ HS has coverage of $\leq 82\%$, decreasing for increasing values of $\varrho$. However, for increasing number of studies (whether large or small), HS appears to converge towards nominal coverage. In particular, for $K=40$ and $\varrho>0.7$ HS provides the most accurate coverage under the beta distribution model. %Complete simulation results for $K \in \{5,10,20,40\}$, $\bar{n} \in \{20,80\}$ and $\tau \in \{0,0.16,0.4\}$ for both models are provided in the supplement in Figures \ref{pp_normal_0}--\ref{pp_beta_4}.

\subsubsection{Interval Lengths}
We simulated the expected confidence interval lengths for all methods discussed in this paper. The detailed results are provided in Figures \ref{pp_length_normal_0} -- \ref{pp_length_beta_4} in the supplement. The results again depend on both the assumed model and the amount of heterogeneity $\tau$.

Generally we observe that the confidence intervals become increasingly narrow for increasing values of $\rho$ and increasingly wide for larger values of $\tau$. For the truncated normal distribution model and $\tau=0$, HS (on average) yields the shortest confidence intervals and HOVz the widest, with the other CIs lying in between with quite similar lengths. Only for $K=5$ the CIs based on the wild bootstrap are quite wide, indicating that potentially more studies are required to reliably use the wild bootstrap based approaches. For $\tau=0.16$ HS again yields the shortest CIs in all scenarios. For small K, the WBS approaches yield the widest CIs and for more studies, HOVz is the widest, when $\rho$ is small, but becoming nearly as narrow as HS when $\rho$ is close to 1. The lengths of the other CIs are nearly identical for $K=40$, whereas for fewer studies there are considerable differences. This relative evaluation also holds for $\tau=0.4$.

When the underlying model is the beta distribution model and $\tau=0$, the results are equivalent to the truncated normal distribution model. For $\tau=0.16$ and $K=5$ the widths of the new CIs decrease with increasing $\varrho$ until $\rho=0.7$. Interestingly, the widths of these CIs then increase again for $\rho>0.7$, which could not be observed in the truncated normal model. This effect becomes much less pronounced for increasing number of studies K. HS is always more narrow than the new CIs and for $K \geq 20$ HOVz is the widest at $\rho=0$ but even more narrow than HS for $\rho \geq 0.8$. For $\tau=0.4$ the results are similar, except that the widths of the CIs now decrease monotonously for increasing $\rho$ and HOVz is most narrow for $\rho > 0.5$.

\subsubsection{Recommendations}
\label{Recommend}
We summarize our findings  by providing recommendations to practitioners wishing to choose between the considered methods. The recommendations will depend on the assumed model and how much heterogeneity is present in the data. We believe the beta distribution model is better suited for random-effects meta-analyses of correlations. Reminder: HOVz employs the inverse Fisher transformation, whereas our newly proposed confidence intervals employ the integral z-to-r transformation suggested by \cite{hafdahl2009improved}.

\begin{itemize}
\item $\tau=0$ \textbf{(Fixed-Effect Model)}: HS and HOVz are not recommendable. We recommend using KH, HC3 or HC4.
\item $\tau=0.16$: \textbf{Truncated normal model:} HS and HOVz are not recommendable and we recommend using KH, HC3 or HC4. For $|\rho| > 0.7$, all methods are unsatisfactory and only in case of $K=40$, HOVz may be preferable. \textbf{Beta distribution model:} HS and HOVz are not recommendable. All new confidence intervals exhibit satisfactory coverage. For small K, WBS approaches yield wider confidence intervals, therefore preferably use KH, HC3 or HC4.

\item $\tau=0.4$: \textbf{Truncated normal model:} HS is not recommendable. For $K=5$ and $|\rho| \leq 0.7$ we again recommend KH, HC3 or HC4. For $K \geq 10$ and $|\rho| \leq 0.7$ we recommend HOVz. For $|\rho| > 0.7$ none of the methods are satisfactory. \textbf{Beta distribution model:} HOVz is not recommendable. For $|\rho| \leq 0.7$ we recommend KH, HC3 or HC4. For $K \geq 40$ and $|\rho| > 0.7$ we recommend using HS. For $K \leq 20$ and $|\rho| > 0.7$ none of the methods are satisfactory.
\end{itemize}

\section{Illustrative Data Analyses}
\label{DataExampleSec}

%We now consider the previously mentioned dataset from \citep{molloy2013conscientiousness} in more detail.

Between 25 and $50\%$ of patients fail to take their medication as prescribed by their caretaker  \citep{molloy2013conscientiousness}. Some studies have shown that medication adherence tends to be better in patients who score higher in conscientiousness (from the five-factor model of personality). Table \ref{molloy} contains data on 16 studies, which investigated the correlation between conscientiousness and medication adherence. These studies were first analyzed in the form of a meta-analysis in \cite{molloy2013conscientiousness}. The columns of Table \ref{molloy} contain information on the authors of the respective study, the year of publication, the sample size of study $i$ ($n_i$), the observed correlation in study $i$, the number of variables controlled for (controls), study design, the type of adherence measure (a\_measure), the type of conscientiousness measure (c\_measure), the mean age of study participants (mean\_age) and the methodological quality (as scored by the authors on a scale from one to four, with higher scores indicating higher quality).

Regarding the measurement of conscientiousness: Where NEO (\textit{Neuroticism-Extraversion-Openness}) is indicated as c\_measure, the personality trait of conscientiousness was measured by one of the various types of NEO personality inventories \citep[PI][]{costa1985neo, costa2008revised}.

\vspace{0.3cm}
\begin{table}[H]
\caption{Data from 16 studies investigating the correlation between conscientiousness and medication adherence}\label{molloy}
\centering
\resizebox{\columnwidth}{!}{
\begin{tabular}{clrrrllllrr}
\toprule
Study $i$ & authors & year & n$_i$ & $r_i$ & controls & design & a\_measure & c\_measure & mean\_age & quality \\ 
\midrule
1 & Axelsson et al. & 2009 & 109 & 0.19 & none & cross-sectional & self-report & other & 22.00 &   1 \\ 
2 & Axelsson et al. & 2011 & 749 & 0.16 & none & cross-sectional & self-report & NEO & 53.59 &   1 \\ 
3 & Bruce et al. & 2010 &  55 & 0.34 & none & prospective & other & NEO & 43.36 &   2 \\ 
4 & Christensen et al. & 1999 & 107 & 0.32 & none & cross-sectional & self-report & other & 41.70 &   1 \\ 
5 & Christensen \& Smith & 1995 &  72 & 0.27 & none & prospective & other & NEO & 46.39 &   2 \\ 
6 & Cohen et al. & 2004 &  65 & 0.00 & none & prospective & other & NEO & 41.20 &   2 \\ 
7 & Dobbels et al. & 2005 & 174 & 0.17 & none & cross-sectional & self-report & NEO & 52.30 &   1 \\ 
8 & Ediger et al. & 2007 & 326 & 0.05 & multiple & prospective & self-report & NEO & 41.00 &   3 \\ 
9 & Insel et al. & 2006 &  58 & 0.26 & none & prospective & other & other & 77.00 &   2 \\ 
10 & Jerant et al. & 2011 & 771 & 0.01 & multiple & prospective & other & NEO & 78.60 &   3 \\ 
11 & Moran et al. & 1997 &  56 & -0.09 & multiple & prospective & other & NEO & 57.20 &   2 \\ 
12 & O'Cleirigh et al. & 2007 &  91 & 0.37 & none & prospective & self-report & NEO & 37.90 &   2 \\ 
13 & Penedo et al. & 2003 & 116 & 0.00 & none & cross-sectional & self-report & NEO & 39.20 &   1 \\ 
14 & Quine et al. & 2012 & 537 & 0.15 & none & prospective & self-report & other & 69.00 &   2 \\ 
15 & Stilley et al. & 2004 & 158 & 0.24 & none & prospective & other & NEO & 46.20 &   3 \\ 
16 & Wiebe \& Christensen & 1997 &  65 & 0.04 & none & prospective & other & NEO & 56.00 &   1 \\ 
\bottomrule
\end{tabular}
}
\end{table}

We performed both a fixed- and random-effects meta-analysis, using all considered methods. For the random-effects model we used the SJ estimator to estimate the between-study heterogeneity variance $\tau^2$.
Combining all available studies yielded $r_{FE} = 0.130$,  $r_{RE} = 0.154$ and $\hat\tau^2_{SJ}= 0.012$. In addition to a complete-case study, we also examined the cross-sectional and prospective studies separately. In total there were five cross-sectional and eleven prospective studies in the dataset. For the cross-sectional studies $r_{FE} = 0.168$ and $r_{RE} = 0.170$ resulted and slightly lower values for the prospective studies ($r_{FE} = 0.108$,  $r_{RE} = 0.147$). Heterogeneity estimates were $\hat{\tau}^2_{SJ} = 0.007$ (cross-sectional) and $\hat{\tau}^2_{SJ} =0.016$ (prospective), respectively. In Table \ref{dataex_ci} we provide values of all CIs discussed in this paper.

\begin{table}[H]
\caption{Random-effects model confidence intervals for all studies and subgroups separated by study design, original data from \cite{molloy2013conscientiousness}}
\label{dataex_ci}
\renewcommand{\arraystretch}{1.4}
\centering
%\resizebox{\columnwidth}{!}{
\begin{tabular}{rcccc}
\toprule
 & & Study design\\
\cmidrule(l){2-4} 
Approach & All Designs & cross-sectional & prospective\\
\midrule
HOVz  & [0.081, 0.221] & [0.067, 0.266] & [0.050, 0.240] \\
HS  & [0.073, 0.174] & [0.100, 0.220] & [0.035, 0.166] \\
KH  & [0.080, 0.218] & [0.037, 0.291] & [0.043, 0.239] \\
\midrule
WBS1  & [0.086, 0.213] & [0.063, 0.267] & [0.051, 0.232]\\
WBS2  & [0.079, 0.219] & [0.053, 0.276] & [0.043, 0.239]\\
WBS3  & [0.084, 0.215] & [0.058, 0.272] & [0.048, 0.234]\\
\midrule
HC3  & [0.081, 0.218] & [0.041, 0.288] & [0.041, 0.241] \\
HC4  & [0.083, 0.216] & [0.054, 0.276] & [0.045, 0.237] \\
\bottomrule
\end{tabular}
\end{table}

%The CIs based on the Fisher transformation are quite similar, with HOVz being just slightly wider than the new CIs. The HS CI is much narrower, which could also be observed in our simulations. However, that the sacrifice for this narrow CI is poor coverage.

In the case of all studies ($K=16$), all methods yield quite similar CIs except for HS. Additional simulations for this situation ($K=16$, $\tau^2 = 0.012$, $n_i$ as in Table \ref{dataex_ci}) are given in the supplement and show a coverage of around 80\% for HS, while all other methods exhibit a fairly accurate coverage of around 95\% and HOVz with around 94\%. Thus, the sacrifice for the narrow HS CIs is poor coverage. Additional analyses of other datasets are given in the supplement.

\section{Discussion}

We introduced several new methods to construct confidence intervals of the main effect in random-effects meta-analyses of correlations, based on the Fisher-z transformation. We compared these to the standard HOVz and Hunter-Schmidt confidence intervals and, following the suggestion by \cite{hafdahl2009improved}, utilized an integral z-to-r transformation instead of the inverse Fisher transformation. We performed an extensive Monte Carlo simulation study, in order to assess the coverage and mean interval length of all CIs.  In addition to the truncated normal distribution model considered by \cite{hafdahl2009meta} and \cite{field2005meta} we also investigated a transformed beta distribution model, which exhibits less bias in the generation of the study level effects.

The results of our simulations show that for low and moderate heterogeneity and correlations of $|\varrho| \leq 0.7$, our newly proposed confidence intervals improved coverage considerably over the classical HOVz and Hunter-Schmidt approaches. However, for extreme heterogeneity and $|\varrho| > 0.7$ all confidence intervals performed poorly. Therefore, further methodological research is necessary in order to fill this gap. Also, the choice of data-generating model (truncated normal or transformed beta distribution) has substantial influence on results. Due to various aspects, which we discussed when introducing the two models, the beta distribution model is arguably more appropriate. Based on our findings, we provide recommendations to practitioners looking for guidance in choosing a method for data analysis. These are listed in subsection \ref{Recommend}.

We attempted to further improve the proposed confidence intervals with the help of a bias correction for the Pearson correlation coefficient $r$,  given by $r^* = \frac{r(1-r^2)}{2(n-1)}$, as the (negative) bias of $r$ is usually approximated by $\mathcal{B}_r = -\frac{\varrho(1-\varrho^2)}{2(n-1)}$ \citep{hotelling1953new,schulze2004meta}. However, this bias correction actually made coverage worse in the studied settings.

\newpage
\bibliographystyle{apa}
\bibliography{korr_lit}

\newpage
\section*{Acknowledgments}

The authors gratefully acknowledge the computing time provided on the Linux HPC cluster at Technical University Dortmund (LiDO3), partially funded in the course of the Large-Scale Equipment Initiative by the German Research Foundation (DFG) as project 271512359. Furthermore, we thank Marl\'ene Baumeister and Lena Schmid for many helpful discussions, and Philip Buczak for finding interesting data sets. This work was supported by the German Research Foundation project (Grant no. PA-2409 7-1).

\section*{Data Availability Statement}
The R-scripts used for our simulations and data analyses will be made publicly available on \texttt{figshare} (pending publication). The dataset from \cite{molloy2013conscientiousness} can be found in the \texttt{metafor} package in \texttt{R} and the datasets considered for re-analysis are from \cite{chalkidou2012correlation} and \cite{santos2016role} respectively.

\newpage
\appendix
\section*{Supplement}

\section{Complete Results of Simulation Study}

We present the complete simulation results regarding coverage and interval lengths for both models under the settings $K \in \{5,10,20,40\}$ and for mean study sizes $\bar{n} \in \{20,80\}$. Additionally we considered the RMSE of the variance estimates of $\bar{z}$ for the confidence intervals based on the Fisher-z transformation.

\begin{figure}[H]
\centering
\includegraphics[width=\textwidth]{Grafiken/normal_coverage_0.pdf}
\caption{Mean Coverage for truncated normal distribution model with $\tau = 0$}
\label{pp_normal_0}
\end{figure}

\begin{figure}[H]
\centering
\includegraphics[width=\textwidth]{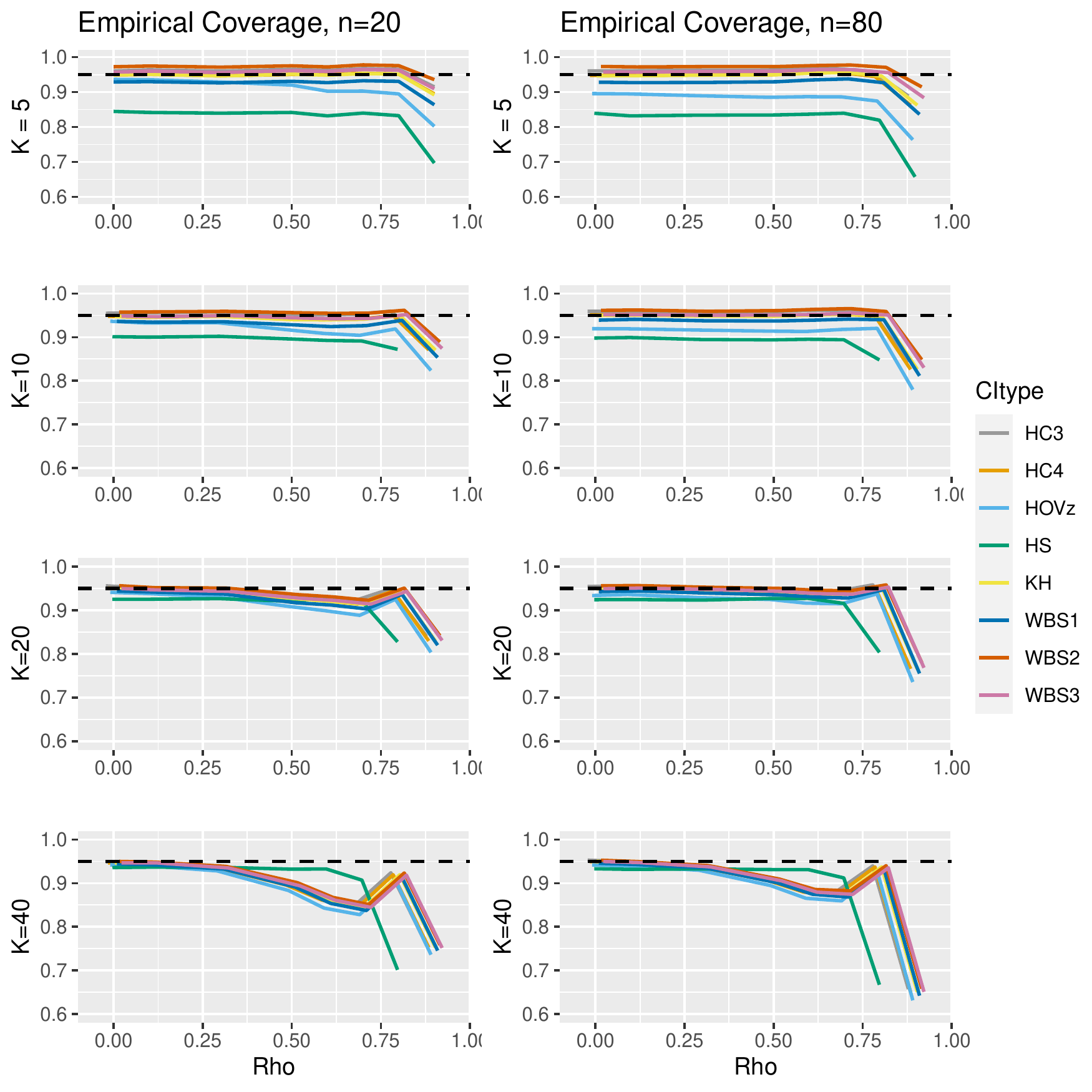}
\caption{Mean Coverage for truncated normal distribution model with $\tau = 0.16$}
\label{pp_normal_16}
\end{figure}

\begin{figure}[H]
\centering
\includegraphics[width=\textwidth]{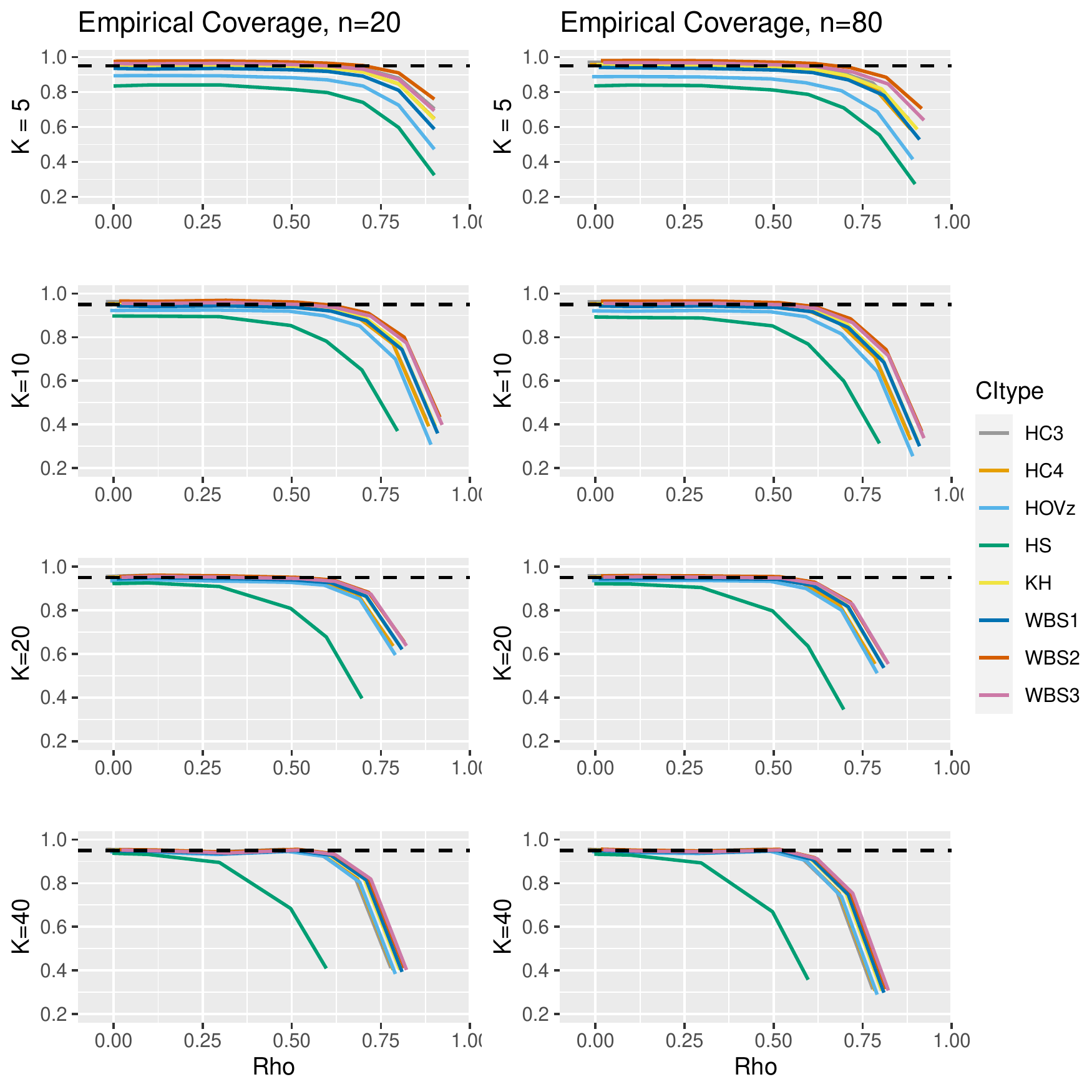}
\caption{Mean Coverage for truncated normal distribution model with $\tau = 0.4$}
\label{pp_normal_4}
\end{figure}

\begin{figure}[H]
\centering
\includegraphics[width=\textwidth]{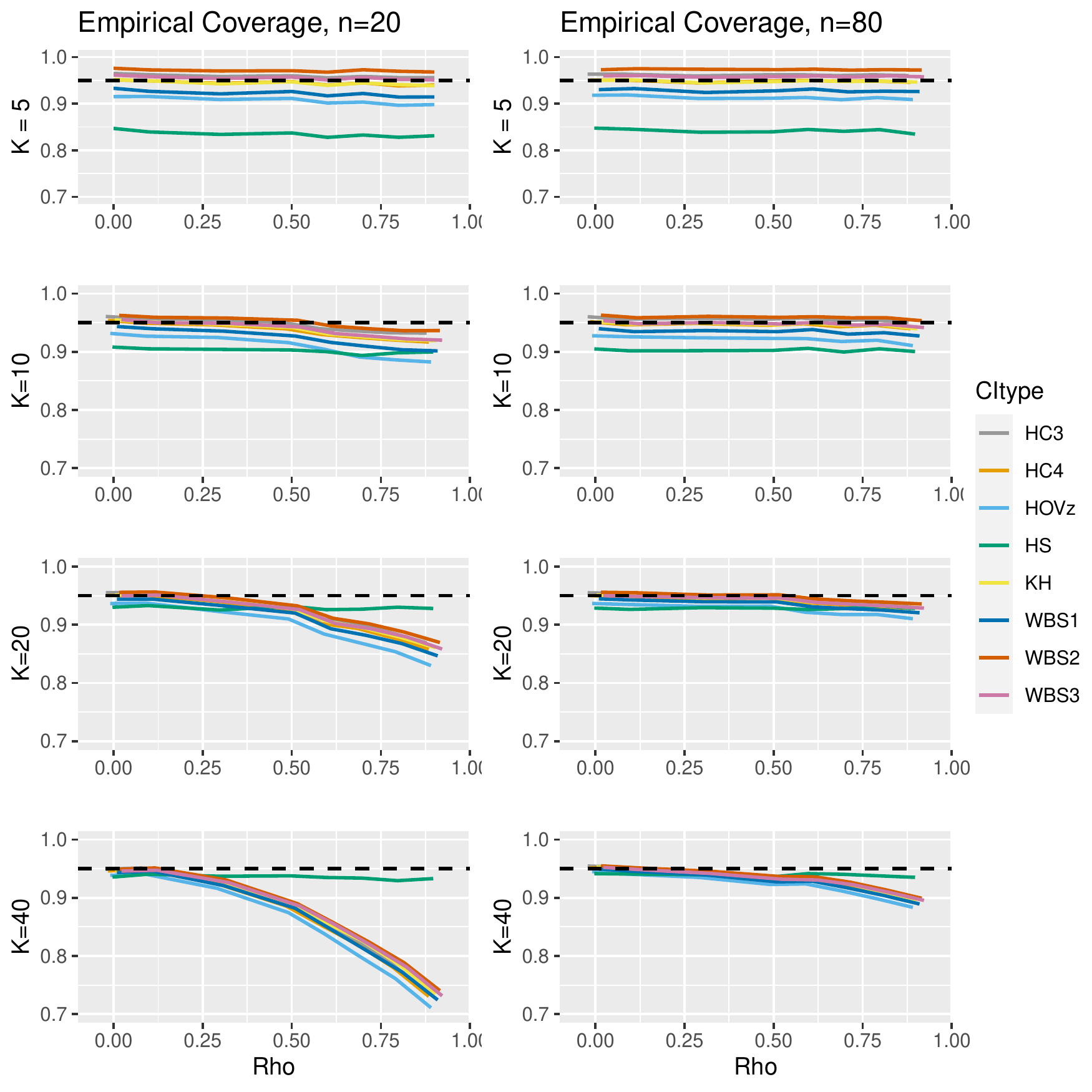}
\caption{Mean Coverage for transformed beta distribution model with $\tau = 0$}
\label{pp_beta_0}
\end{figure}

\begin{figure}[H]
\centering
\includegraphics[width=\textwidth]{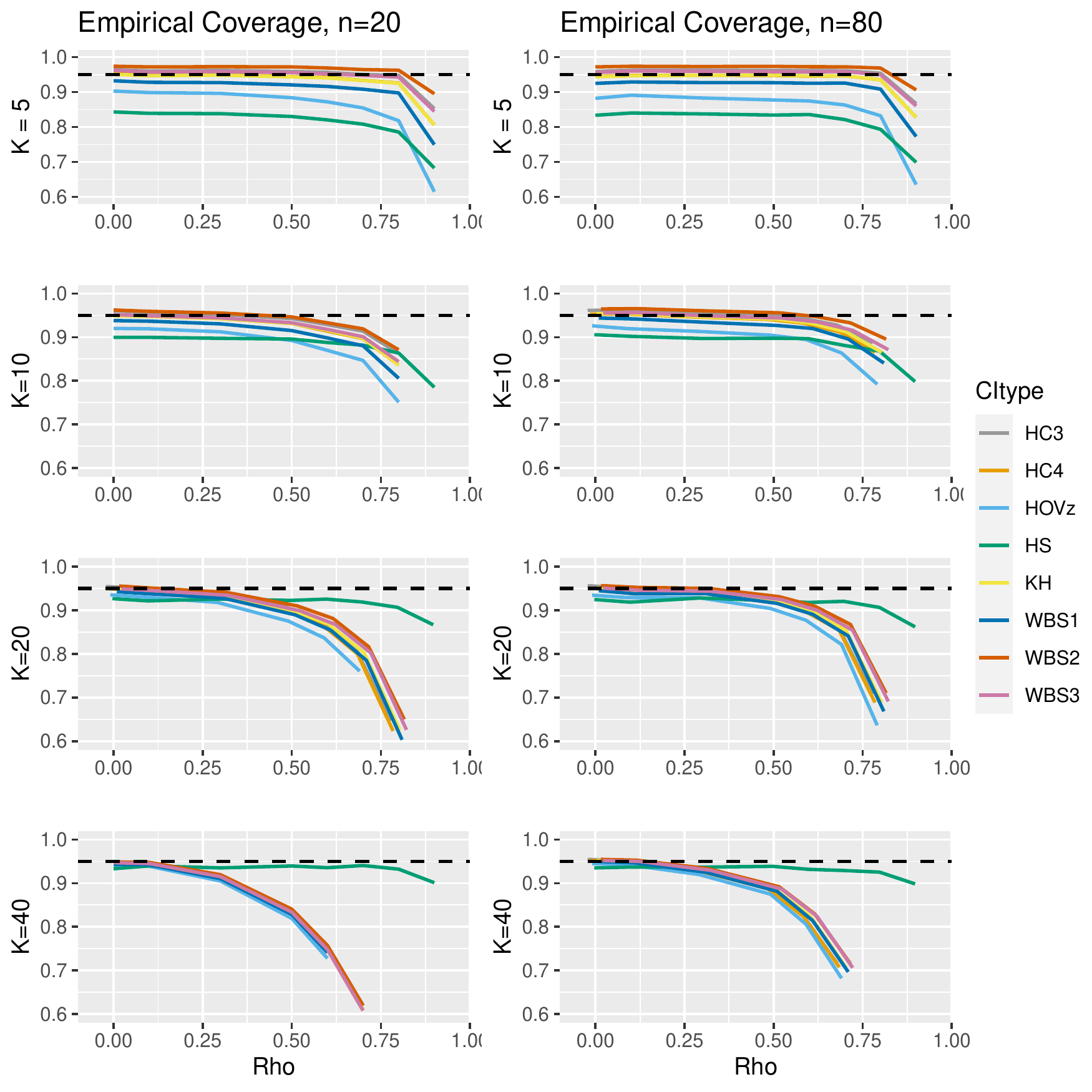}
\caption{Mean Coverage for transformed beta distribution model with $\tau = 0.16$}
\label{pp_beta_16}
\end{figure}

\begin{figure}[H]
\centering
\includegraphics[width=\textwidth]{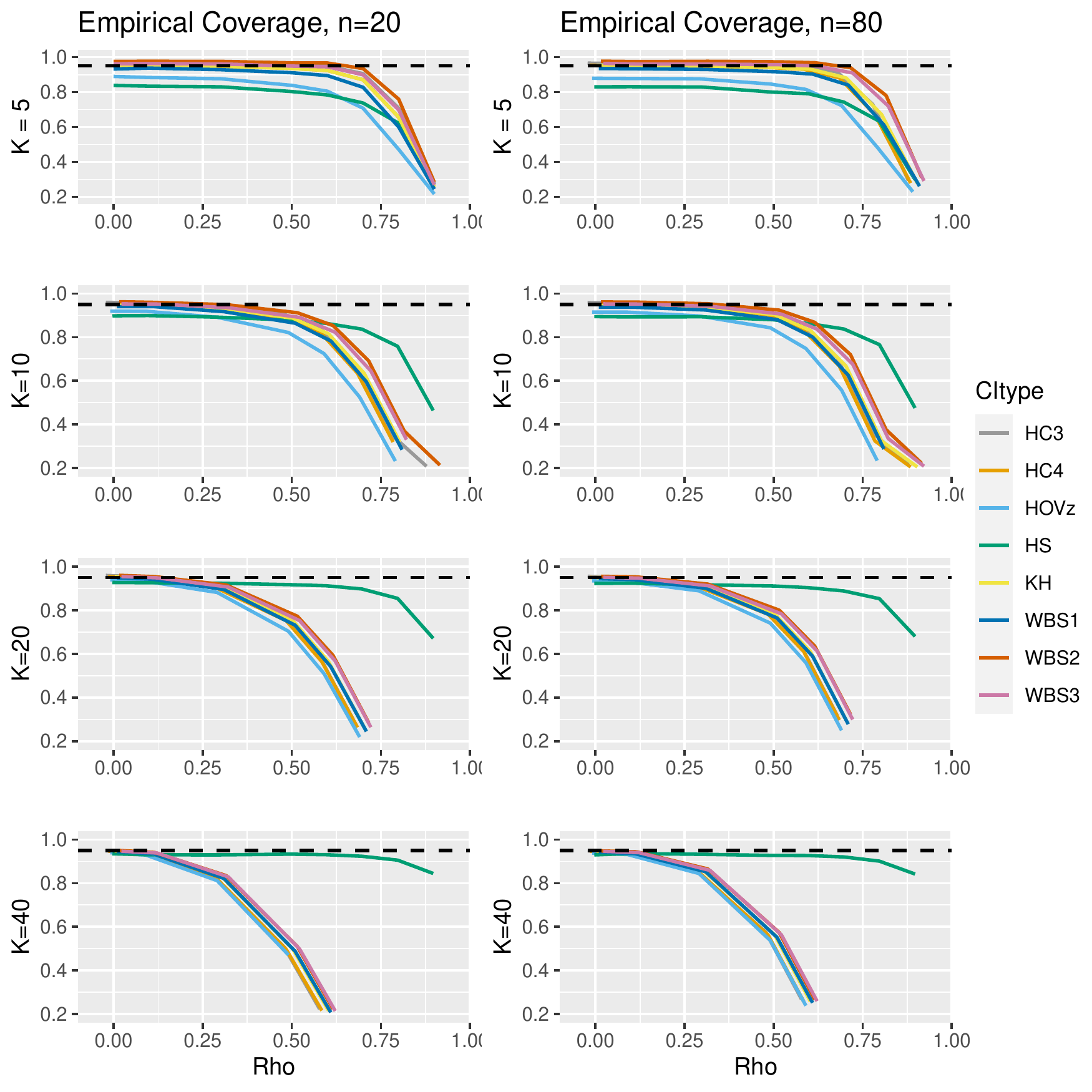}
\caption{Mean Coverage for transformed beta distribution model with $\tau = 0.4$}
\label{pp_beta_4}
\end{figure}

%%%%%%%%%%%%%%%%%%%%%%%%%%%%%%%%%%%%%%%%%%%%%%%%%%%%%%%%%%%%%%%%%%%%%%%%%%%%%%%%%%%
%%%  Interval lengths:

\begin{figure}[H]
\centering
\includegraphics[width=\textwidth]{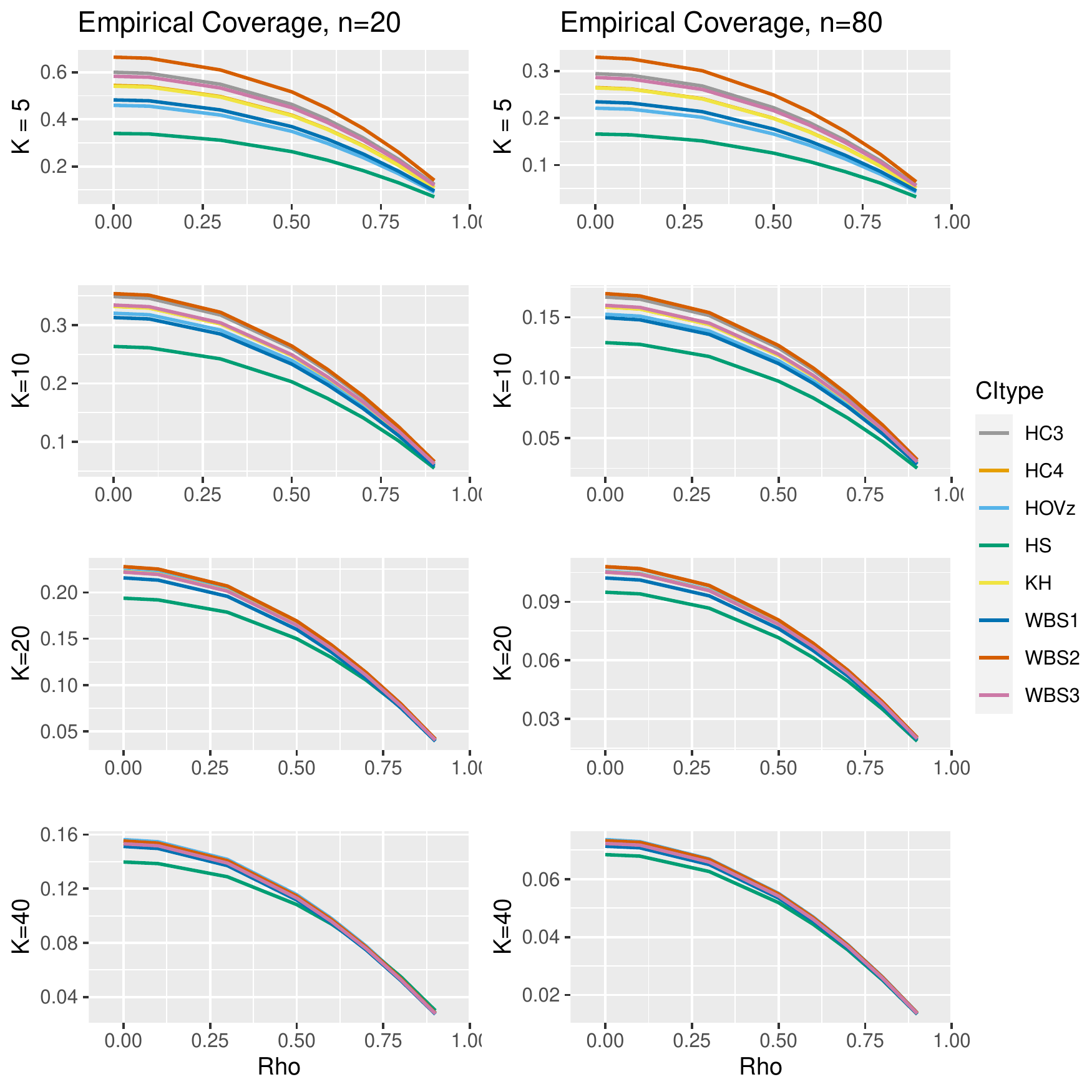}
\caption{Mean CI length for truncated normal distribution model with $\tau = 0$}
\label{pp_length_normal_0}
\end{figure}

\begin{figure}[H]
\centering
\includegraphics[width=\textwidth]{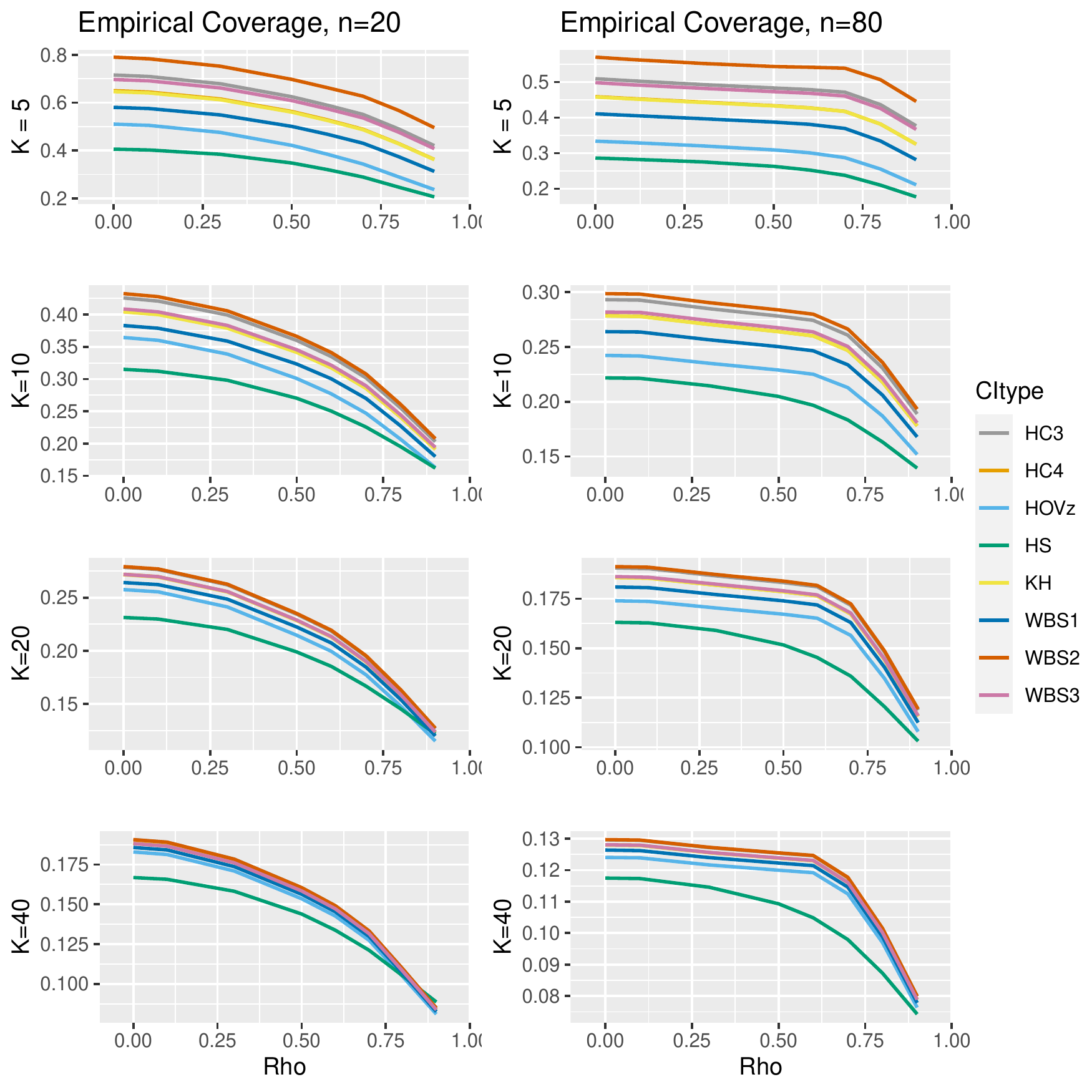}
\caption{Mean CI length for truncated normal distribution model with $\tau = 0.16$}
\label{pp_length_normal_16}
\end{figure}

\begin{figure}[H]
\centering
\includegraphics[width=\textwidth]{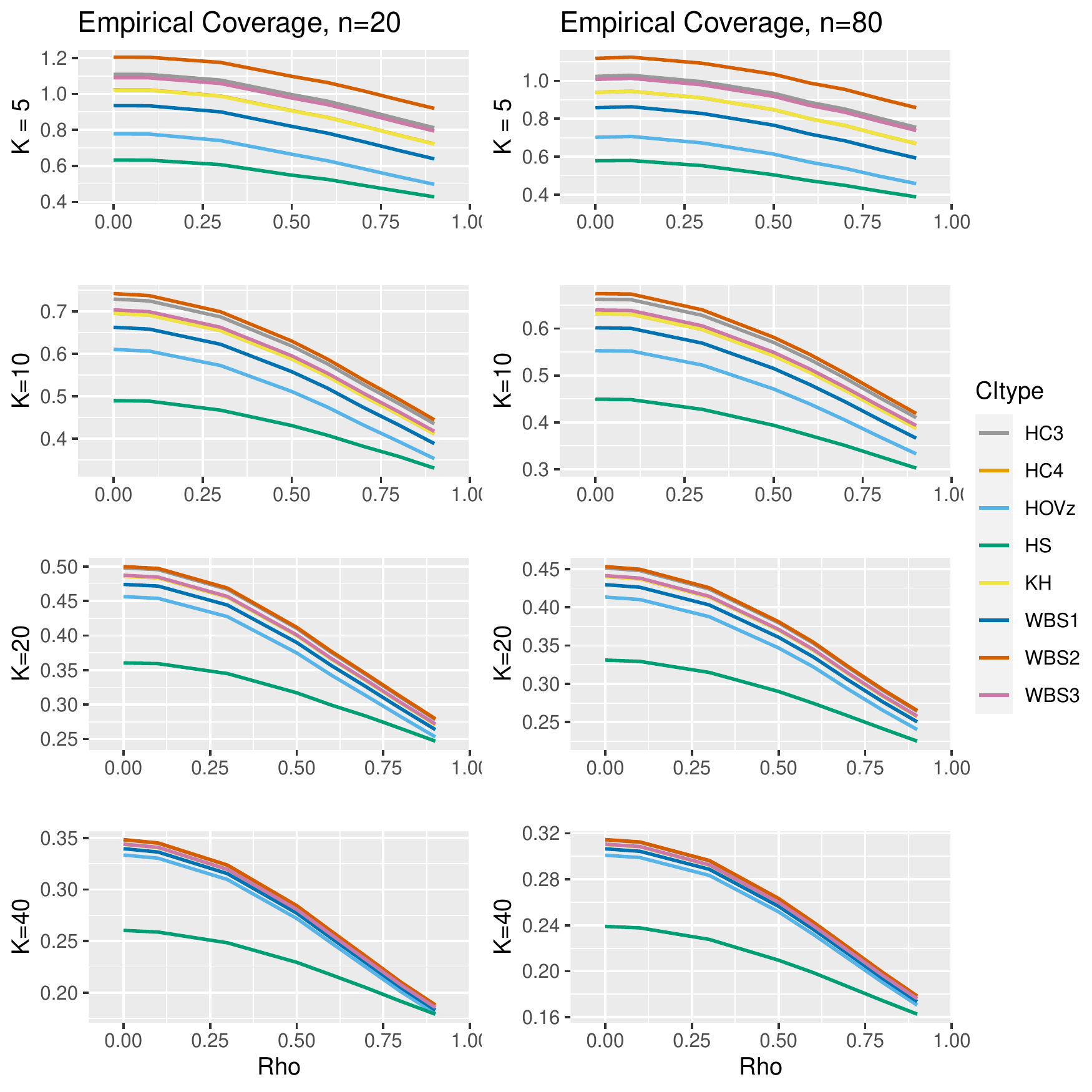}
\caption{Mean CI length for truncated normal distribution model with $\tau = 0.4$}
\label{pp_length_normal_4}
\end{figure}

\begin{figure}[H]
\centering
\includegraphics[width=\textwidth]{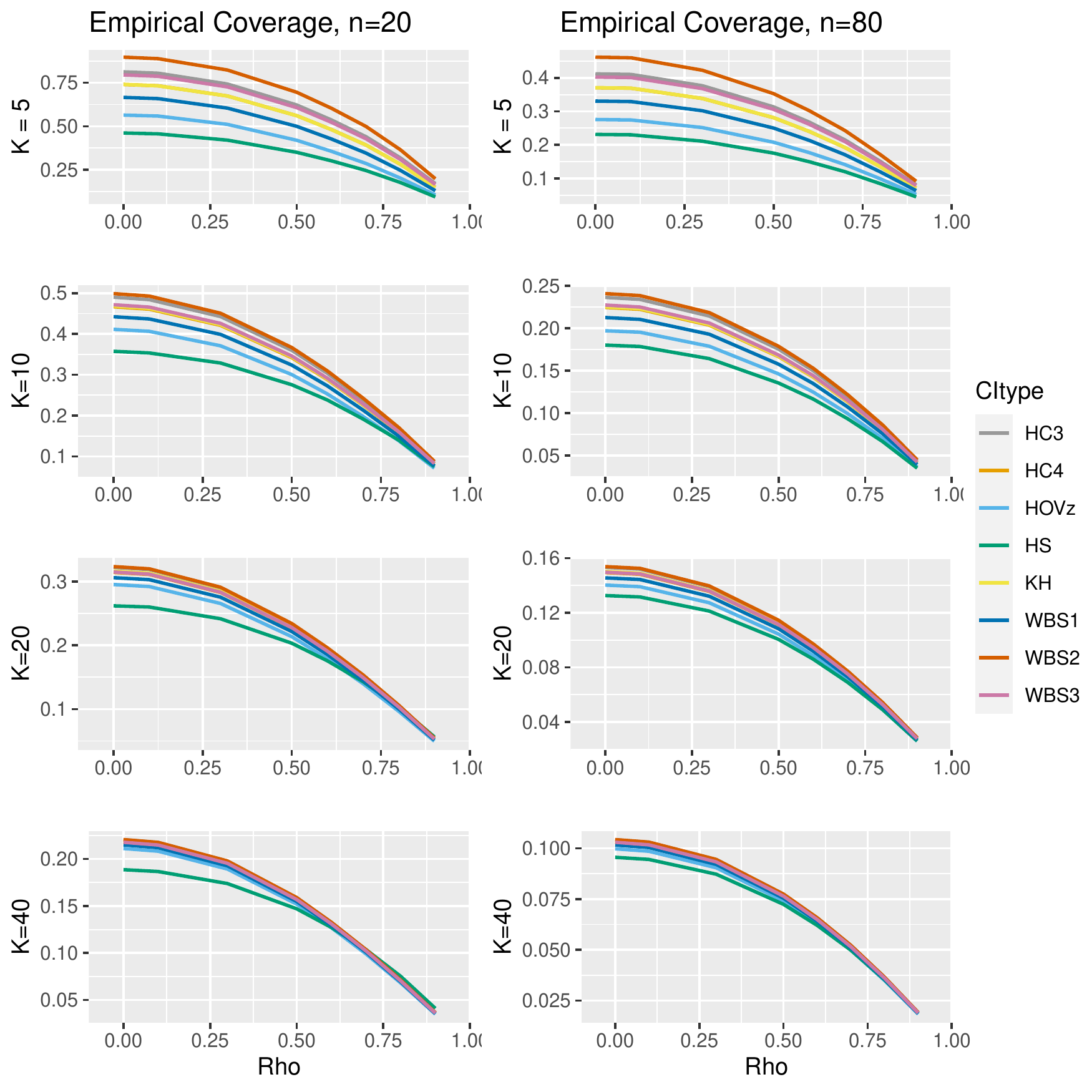}
\caption{Mean CI length for transformed beta distribution model with $\tau = 0$}
\label{pp_length_beta_0}
\end{figure}

\begin{figure}[H]
\centering
\includegraphics[width=\textwidth]{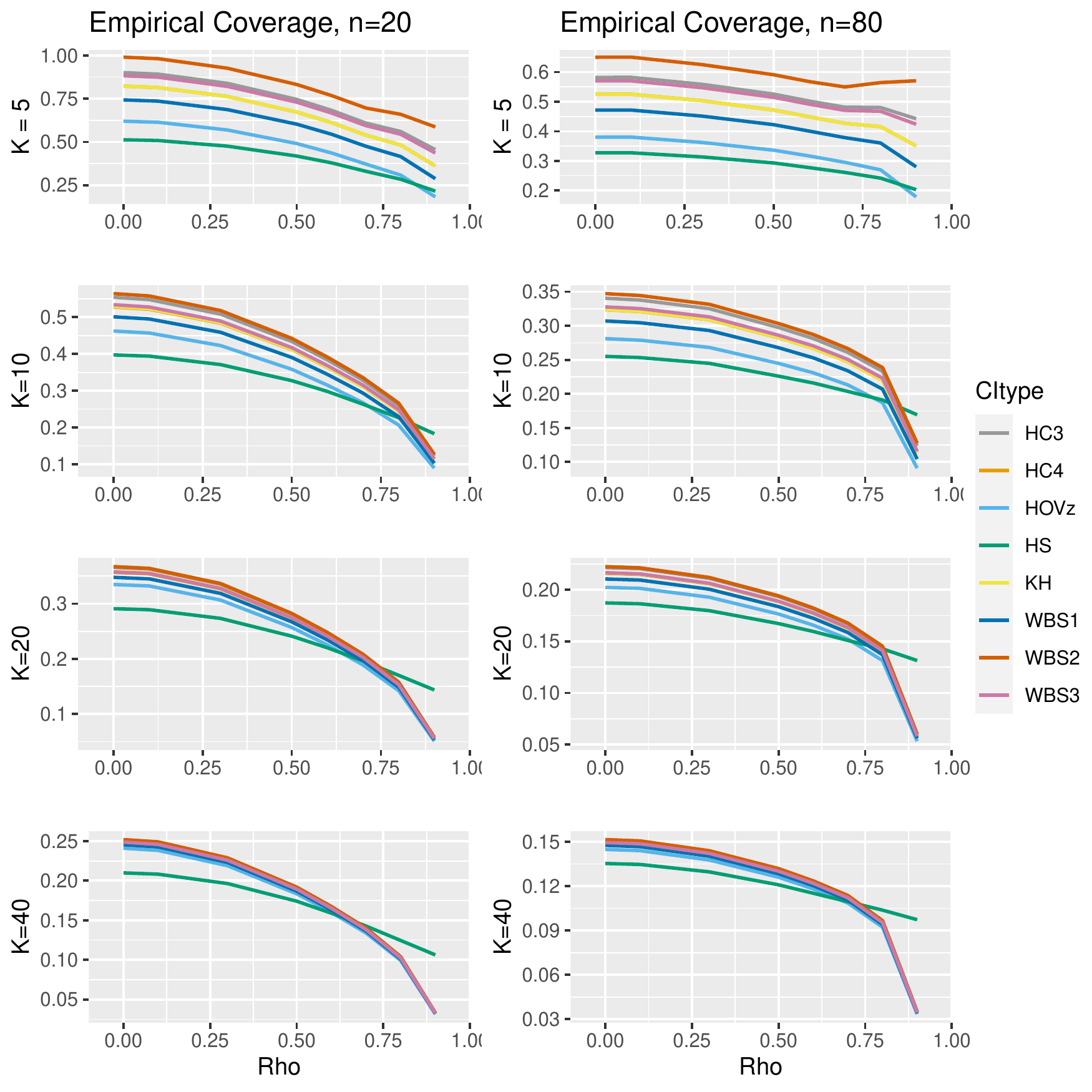}
\caption{Mean CI length for transformed beta distribution model with $\tau = 0.16$}
\label{pp_length_beta_16}
\end{figure}

\begin{figure}[H]
\centering
\includegraphics[width=\textwidth]{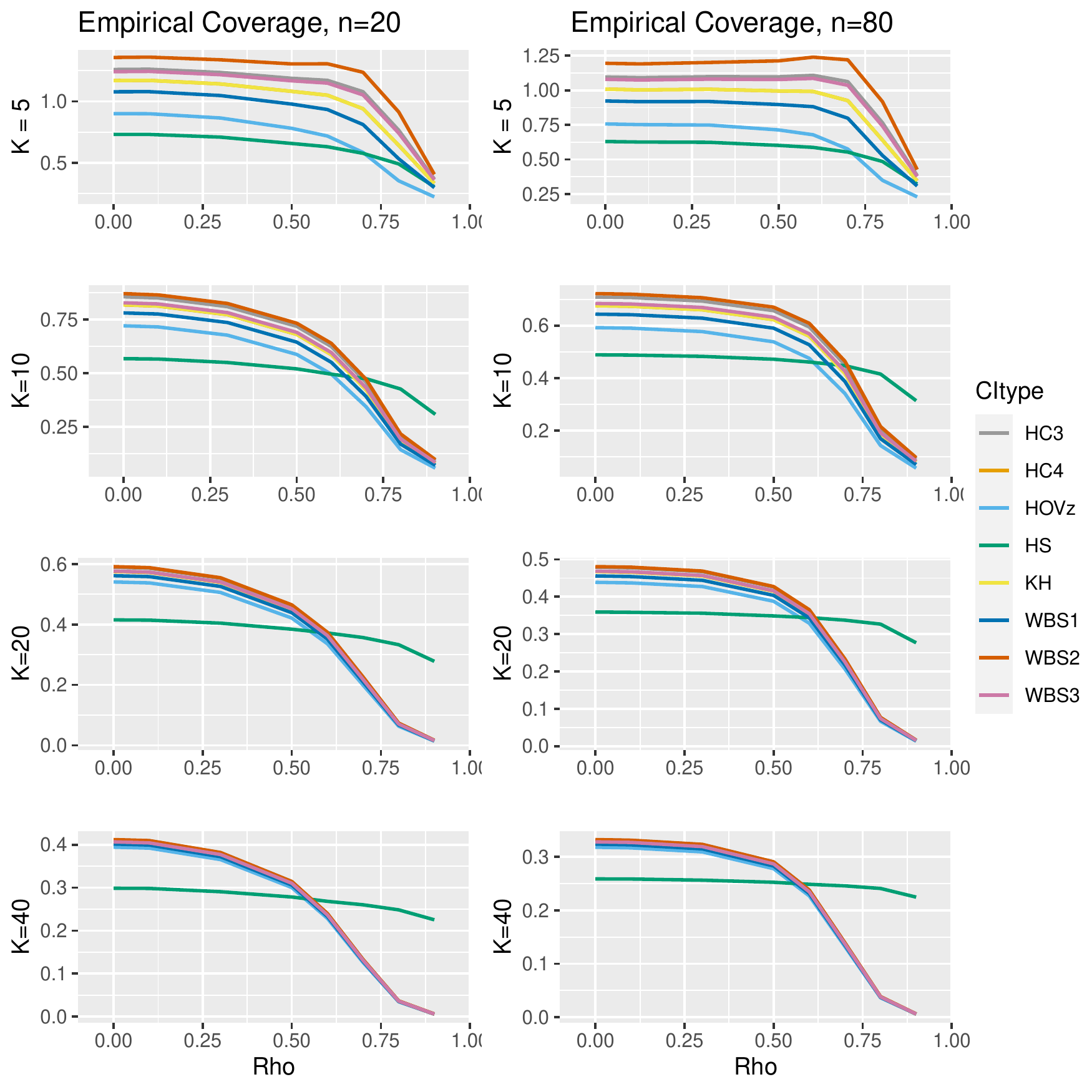}
\caption{Mean CI length for transformed beta distribution model with $\tau = 0.4$}
\label{pp_length_beta_4}
\end{figure}

\section{Simulations based on the dataset from Section \ref{DataExampleSec}}

We also added a simulation setting that is specific to the dataset from \citep{molloy2013conscientiousness} discussed in Section \ref{DataExampleSec}. This means the number of studies, \enquote{true} heterogeneity and study effects in the simulation were chosen according to the estimates from the original dataset. There were $K = 16$ studies, with $\hat{\tau}^2 = 0.012$ and a range of study sizes between 55 and 771. The results are displayed in Table \ref{dataExTable}. Our newly proposed confidence intervals have good control of the nominal coverage $95\%$ both for the truncated normal and beta distribution simulation designs. HOVz was slightly conservative with approximately $94\%$ coverage. HS performed worst out of the considered approaches, with only around $80\%$ coverage.

\begin{table}[H]
\centering
\begin{tabular}{l|cccccccc}
distribution & HOVz & KH & WBS1 & WBS2 & WBS3 & HC3 & HC4 & HS \\ 
\hline
normal & 0.938 & 0.954 & 0.946 & 0.948 & 0.947 & 0.954 & 0.948 & 0.798\\
beta & 0.940 & 0.953 & 0.947 & 0.946 & 0.947 & 0.954 & 0.949 & 0.797\\
\end{tabular}
\caption{Empirical coverage in simulation setting based on data from \cite{molloy2013conscientiousness} with K=16, $\tau^2 = 0.012$ and study sizes between 55 and 771}
\label{dataExTable}
\end{table}

%In Tables \ref{truncnormcomplete} and \ref{betacomplete} we present the complete simulation results for all settings. Table \ref{truncnormcomplete} contains the results for the classical simulations with a truncated normal distribution. Table \ref{betacomplete} shows the results for our new simulation design with a transformed beta distribution. Table \ref{kuerzel} explains the abbreviations used for the design settings with regard to the respective number and size of studies.

%\begin{table}[H]
%\begin{tabular}{c|l}
%Abbreviation & Study sizes \\
%\hline
%A1 & 15, 16, 19, 23, 27\\
%A2 & 60, 64, 76, 92, 108\\
%A3 & $2 \times$ (15, 16, 19, 23, 27)\\
%A4 & $2 \times$ (60, 64, 76, 92, 108)\\
%B1 & $4 \times$ (15, 16, 19, 23, 27)\\
%B2 & $4 \times$ (60, 64, 76, 92, 108)\\
%B3 & $8 \times$ (15, 16, 19, 23, 27)\\
%B4 & $8 \times$ (60, 64, 76, 92, 108)\\
%C & 23, 19, 250, 330, 29\\
%D & $2 \times$ (210, 240, 350, 220, 290, 280, 340, 400, 380, 290)\\
%\end{tabular}
%\caption{Abbreviations used to describe the number and size of studies in Tables \ref{truncnormcomplete} and \ref{betacomplete}}
%\label{kuerzel}
%\end{table}

\section{Additional Information}

\begin{figure}[H]
\centering
\input{Grafiken/normal_4.tex}
\caption{Mean Coverage for truncated normal distribution model with $\tau = 0.4$, aggregated across all number of studies and study size settings}
\label{normal_4}
\end{figure}

\begin{figure}[H]
\centering
\input{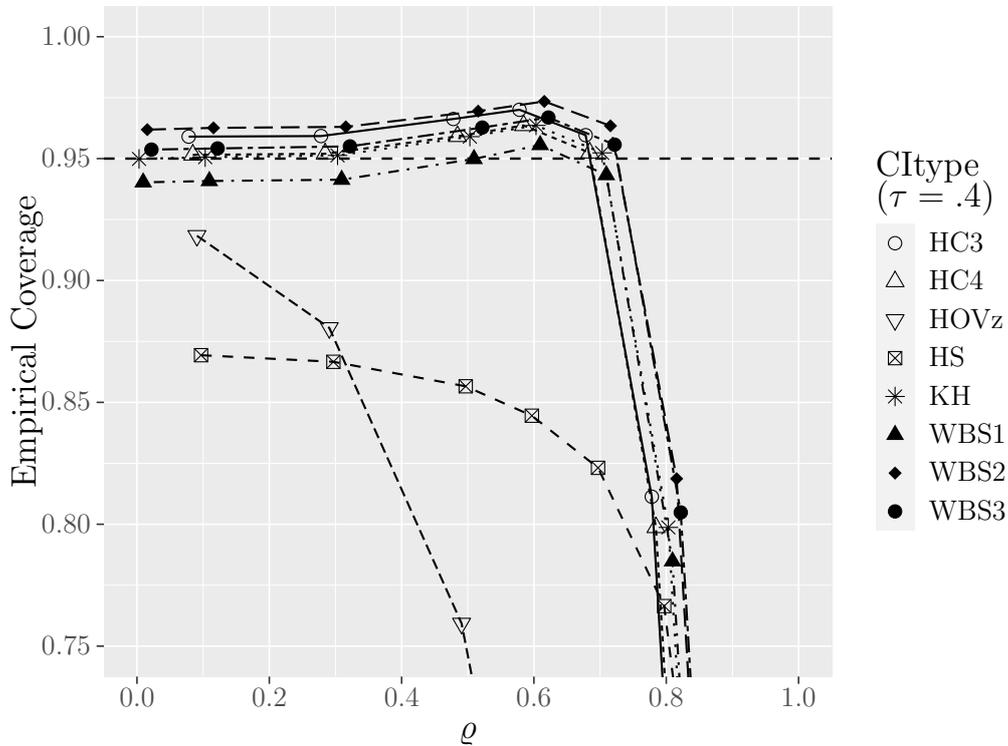}
\caption{Mean Coverage for transformed beta distribution model with $\tau = 0.4$, aggregated across all number of studies and study size settings}
\label{beta_4}
\end{figure}

\underline{Comment regarding Table \ref{mini_simu}:}\\
\vspace{.1cm}

The standardized log-normal distribution simulated in Table \ref{mini_simu}, was generated in the following manner:

\begin{equation*}
Y_i = \frac{X_i - \exp(0.5)}{\sqrt{\exp(2)-\exp(1)}},
\end{equation*}

\noindent
where $X_i \overset{iid}{\sim} \mathcal{LN}(0,1)$. Then the $Y_i$ are iid and follow a standardized log-normal distribution with mean 0 and variance 1.

\begin{figure}[H]
\centering
\includegraphics[width=\textwidth]{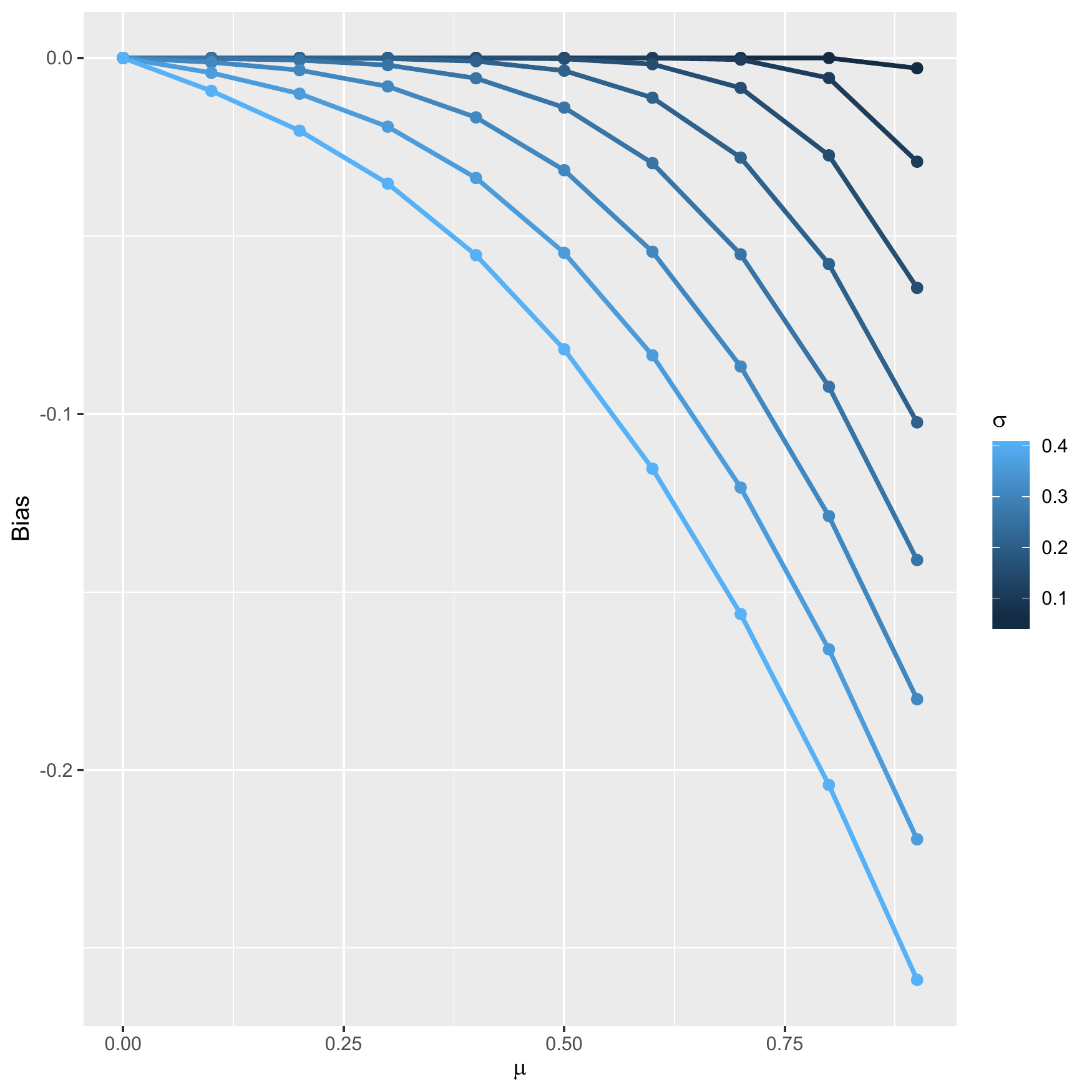}
\caption{Bias of truncated normal distribution on [-0.999,0.999] for various means $\mu$ and standard deviations $\sigma$}
\label{biasplot}
\end{figure}

\section{Reanalysis of other Meta-analyses}

In order to gain additional insights into the consequences of implementing our newly proposed methods in practice, we reanalyzed previous meta-analyses of correlations. To this end we considered two datasets from \cite{chalkidou2012correlation} and \cite{santos2016role}.

\cite{santos2016role} investigated the role of the amygdala in facial trustworthiness through meta-analysis of fMRI studies. They performed a meta-analysis of 12 studies, investigating the correlation between amygdala response to trustworthy vs. untrustworthy facial signals under fMRI. The data is presented in Table \ref{Santos} \footnote[2]{Study number twelve actually reported a correlation of 1, but because we apply the Fisher-z transformation, we truncated this to 0.999.}.

\begin{table}[H]
\centering
\footnotesize{
\begin{tabular}{r|rrrrrrrrrrrr}
Study & 1 & 2 & 3 & 4 & 5 & 6 & 7 & 8 & 9 & 10 & 11 & 12\\
\hline
$r_i$ & .654 & .072 & .998 & .892 & .313 & .069 & -.971 & .989 & .989 & .473 & .594 & .999\\
\hline
$n_i$ & 24 & 16 & 12 & 14 & 15 & 15 & 6 & 12 & 11 & 32 & 14 & 12
\end{tabular}
}
\caption{Reported correlations and sample sizes of 12 studies on amygdala response to facial signals of trustworthiness under fMRI in \cite{santos2016role}.}
\label{Santos}
\end{table}

This is clearly one of the challenging scenarios with extreme correlations and high heterogeneity. For the random-effects meta-analysis \cite{santos2016role} reported a total estimated effect of 0.851 with a $95\%$ confidence interval of $[.422, .969]$. With our new methods (also adding HS) we obtain the following confidence intervals: WBS1: [.088, .764], WBS2: [.044, .785], WBS3: [.066, .775], KH: [.064, .776], HC3: [.050, .782], HC4: [.070, .773], HS: [.302, .784]. Evidently the new CIs are substantially different, with a noticeable shift to smaller values. This makes sense as the integral z-to-r transformation increasingly deviates from the inverse Fisher transform for larger $|\varrho|$ values. Also, as in the simulations, HS yields the most (probably overly) narrow interval.

\cite{chalkidou2012correlation} examined the correlation between ki-67 immunohistochemistry and 18F-Fluorothymidine uptake in patients with cancer. The data comes from a total of 9 studies, containing data from both biopsies and surgeries, and is presented in Table \ref{Chalkidou}.

\begin{table}[H]
\centering
\footnotesize{
\begin{tabular}{r|rrrrrrrrr}
Study & 1 & 2 & 3 & 4 & 5 & 6 & 7 & 8 & 9 \\
\hline
$r_i$ & .21 & .79 & .82 & .80 & .04 & .92 & .84 & .77 & .57 \\
\hline
$n_i$ & 43 & 12 & 9 & 10 & 20 & 20 & 21 & 6 & 22
\end{tabular}
}
\caption{Reported correlations and sample sizes of 9 studies on ki-67 immunohistochemistry and 18F-Fluorothymidine in \cite{chalkidou2012correlation}.}
\label{Chalkidou}
\end{table}

The authors report a random-effects meta-analysis 95\% confidence interval of the main effect of [.43, .86]. With our new methods (also adding HS) we obtain the following confidence intervals: WBS1: [.36, .81], WBS2: [.31, .83], WBS3: [.34, .82], KH: [.36, .81], HC3: [.31, .83], HC4: [.33, .82], HS: [.33, .75]. In this example our results are much closer to the authors' analysis, suggesting a slightly wider confidence interval, mainly due to a smaller lower bound.

These examples show that in real world datasets the confidence intervals obtained through our new methods can both deviate substantially or be quite similar to classical approaches, depending on the specific circumstances like number of studies, amount of heterogeneity, study- and effect sizes.

%%%%%%%%%%%%%%%%%%%%%%%%%%%%%%%%%%%%%%%%%%%%%%%%%%%%%%%%%%%%%%%%%%%%%%%%%%%%%%%%%%%%
%%%%%%%%%%%%%%%%%%%%%%%%%%%%%%%%%%%%%%%%%%%%%%%%%%%%%%%%%%%%%%%%%%%%%%%%%%%%%%%%%%%%
%%%%%%%%%%%%%%%%%%%%%%%%%%%%%%%%%%%%%%%%%%%%%%%%%%%%%%%%%%%%%%%%%%%%%%%%%%%%%%%%%%%%

%We define $T \coloneqq \frac{\hat{z} - \tanh^{-1}(\varrho)}{\hat{\sigma}_z}$ and $w \coloneqq \sum\limits_{i=1}^K w_i$. Then it holds for $\mathbb{E}(\hat{z})$ that
%\begin{align*}
%\mathbb{E}[\hat{z}] & = \mathbb{E} \big[ \sum \frac{w_i}{w} \hat{z}_i \big]\\
%& \overset{\hat{\sigma}_i^2 = \sigma_i^2}{=} \sum \frac{w_i}{w} \mathbb{E}[\hat{z}_i]\\
%& = \sum \frac{w_i}{w} \mathbb{E}[\tanh^{-1}(r_i)]\\
%& \overset{Taylor}{\approx} \sum \frac{w_i}{w} \big[ \tanh^{-1}(\varrho) + R_1(\varrho) \big],
%\end{align*}
%%& = \tanh^{-1}(\varrho).
%
%\noindent where $R_1(\varrho) = \mathbb{E}(\tanh^{-1}(\varrho)) - T_1(\varrho)$ is the remainder corresponding to the second-order Taylor approximation. It holds that $|\frac{d^2}{d \varrho^2} \tanh^{-1}(\varrho)| = |\frac{2\varrho}{(1-\varrho^2)^2}| \underset{\varrho \in [-1+\delta,1-\delta]}{\leq} M$ for a constant $M \in \mathbb{R}$. Therefore $|R_1(\varrho)| \leq M \varepsilon_i^2$ and it follows
%
%\begin{align*}
%|\mathbb{E}(\hat{z}) - \tanh^{-1}(\varrho)| & \approx |\mathbb{E}(R_1(\varrho))|\\
%& \hspace*{-0.27cm} \underset{\text{Jensen}}{\leq} \mathbb{E}(|R_1(\varrho)|)\\
%& \leq  \mathbb{E}(|M \varepsilon_i^2)|)\\
%&  = M \sigma_i^2 \underset{\sigma_i^2 \rightarrow 0}{\overset{a.s.}{\longrightarrow}} 0.
%\end{align*}

\end{document}